\documentclass[12pt]{article}
\pdfoutput=1
\usepackage{float}
\usepackage[final]{graphicx}
\usepackage{amssymb,amsmath}
\usepackage{epsfig}
\usepackage{epstopdf}
\usepackage{latexsym}
\usepackage{graphicx}
\usepackage{subfigure}
\usepackage{booktabs}
\usepackage{tkz-euclide}
\usetkzobj{all}

\usepackage{makeidx}
\usepackage{cite}
\usepackage{bm}
\usepackage{geometry}
\usepackage{braket}
\usepackage[margin=20pt,small]{caption}

\usepackage[toc]{appendix}

\usepackage{tikz}
\usetikzlibrary{calc,decorations.markings,arrows.meta,shapes.misc,decorations.pathmorphing,calc,bending}

\usetikzlibrary{arrows.meta,shapes.misc,decorations.pathmorphing,calc,bending}

\tikzset{
  branch point/.style={cross out,draw=black,fill=none,minimum size=2*(#1-\pgflinewidth),inner sep=0pt,outer sep=0pt}, 
  branch point/.default=5
}
\tikzset{
  branch cut/.style={
    decorate,decoration=snake,
    to path={
      (\tikztostart) -- (\tikztotarget) \tikztonodes
    },
    }
  }

\usepackage{color}
\usepackage{datetime}
  
\ifpdf
\fi

\def\cO{{\cal O}}

\def\I{{\cal I}}

\definecolor{cardinal}{rgb}{0.6,0,0}
\definecolor{darkgreen}{rgb}{0,0.5,0}
\definecolor{golden}{rgb}{0.92, 0.7, 0}
\definecolor{midnight}{rgb}{0, 0, 0.5}
\definecolor{darkblue}{rgb}{0.2, 0, 0.8}

\newcommand{\be}{\begin{equation}}
\newcommand{\ee}{\end{equation}}
\newcommand{\bea}{\begin{eqnarray}}
\newcommand{\eea}{\end{eqnarray}}

\begin{document}

\begin{titlepage}

\bigskip
\bigskip
\bigskip
\centerline{\Large \bf Real Time Dynamics from Low Point Correlators in 2d BCFT}
\bigskip
\centerline{{\bf Suchetan Das$^1$, Bobby Ezhuthachan$^1$, Arnab Kundu$^2$}}
\bigskip
\bigskip
\bigskip
\centerline{$^1$Ramakrishna Mission Vivekananda Educational and Research Institute,}
\centerline{Belur Math,}
\centerline{Howrah-711202, West Bengal, India.} 
\bigskip
\bigskip
\centerline{$^2$Theory Division,}
\centerline{Saha Institute of Nuclear Physics, HBNI,}
\centerline{1/AF Bidhannagar, Kolkata 700064, India.} 
\bigskip
\bigskip
\bigskip
\centerline{suchetan.das[at]rkmvu.ac.in,  bobby.ezhuthachan[at]rkmvu.ac.in, arnab.kundu[at]saha.ac.in}
\bigskip
\bigskip

\begin{abstract}

\noindent In this article, we demonstrate how a $3$-point correlation function can capture the out-of-time-ordered features of a higher point correlation function, in the context of a conformal field theory (CFT) with a boundary, in two dimensions. Our general analyses of the analytic structures are independent of the details of the CFT and the operators, however, to demonstrate a Lyapunov growth we focus on the Virasoro identity block in large-c CFT's. Motivated by this, we also show that the phenomenon of pole-skipping is present in a $2$-point correlation function in a two-dimensional CFT with a boundary. This pole-skipping is related, by an analytic continuation, to the maximal Lyapunov exponent for maximally chaotic systems.   Our results hint that, the dynamical content of higher point correlation functions, in certain cases, may be encrypted within low-point correlation functions, and analytic properties thereof.

\end{abstract}

\newpage

\tableofcontents

\end{titlepage}

\newpage

\rule{\textwidth}{.5pt}\\

\section{Introduction}

Boundaries can play a crucial role in governing the dynamics of a physical system, both in the classical as well as in the quantum mechanical regime. For example, while a free particle is integrable, imposing a set of appropriate boundary conditions, such as the billiard board, can render the dynamics ergodic. Although it seems plausible, to the best of our knowledge, an analogous statement in the quantum mechanical regime is not known.  

The notion of ergodicity in the quantum regime is a subtle issue. In this article, we will adopt a simple and precise notion, following the recent surge of activities in studies of black holes, generic spin systems and conformal field theories, see {\it e.g.}~\cite{Shenker:2013pqa, Shenker:2013yza, Shenker:2014cwa, Maldacena:2015waa, Roberts:2014ifa, Fitzpatrick:2016thx}. For a given system which has a semi-classical regime, {\it i.e.}~the Planck constant $\hbar$ (or, the like of it)\footnote{For example, for systems with many degrees of freedom, $N$, in the $N\to\infty$ limit, a natural semi-classical description holds.} gives rise to a natural hierarchy of time-scales, ergodicity can be defined in terms of real time behaviour of the thermal correlation functions of the system.

Given $n$-point correlators, there are two broad categories in which they can be grouped: Time-ordered (TO) and Out-of-time-ordered (OTO). While the former, as the name suggests, come with a monotonic ordering of time arguments in the operator ordering of the correlator, the latter does not have such a monotonicity. For example, any $2$-point correlator is time-ordered, since $\langle \cO_1 (t_1) \cO_2(t_2) \rangle$ can only have either $t_1 > t_2$ or $t_2 > t_1$. A $3$-point correlator, however, is not so: {\it e.g.}~consider the correlators $\langle \cO_1 (t_1) \cO_2(t_2) \cO_3(t_3)\rangle$ and $\langle \cO_2 (t_2) \cO_1(t_1) \cO_3(t_3)\rangle$, with $t_1 > t_2 > t_3$. Thus, in this case, one has either a time-ordered correlator (TOC) or an out-of-time-ordered correlator (OTOC). A priori, these two classes of correlators can contain distinct dynamical information of the corresponding system. 

\begin{figure}[ht!]
\begin{center}
{\includegraphics[width=0.8\textwidth]{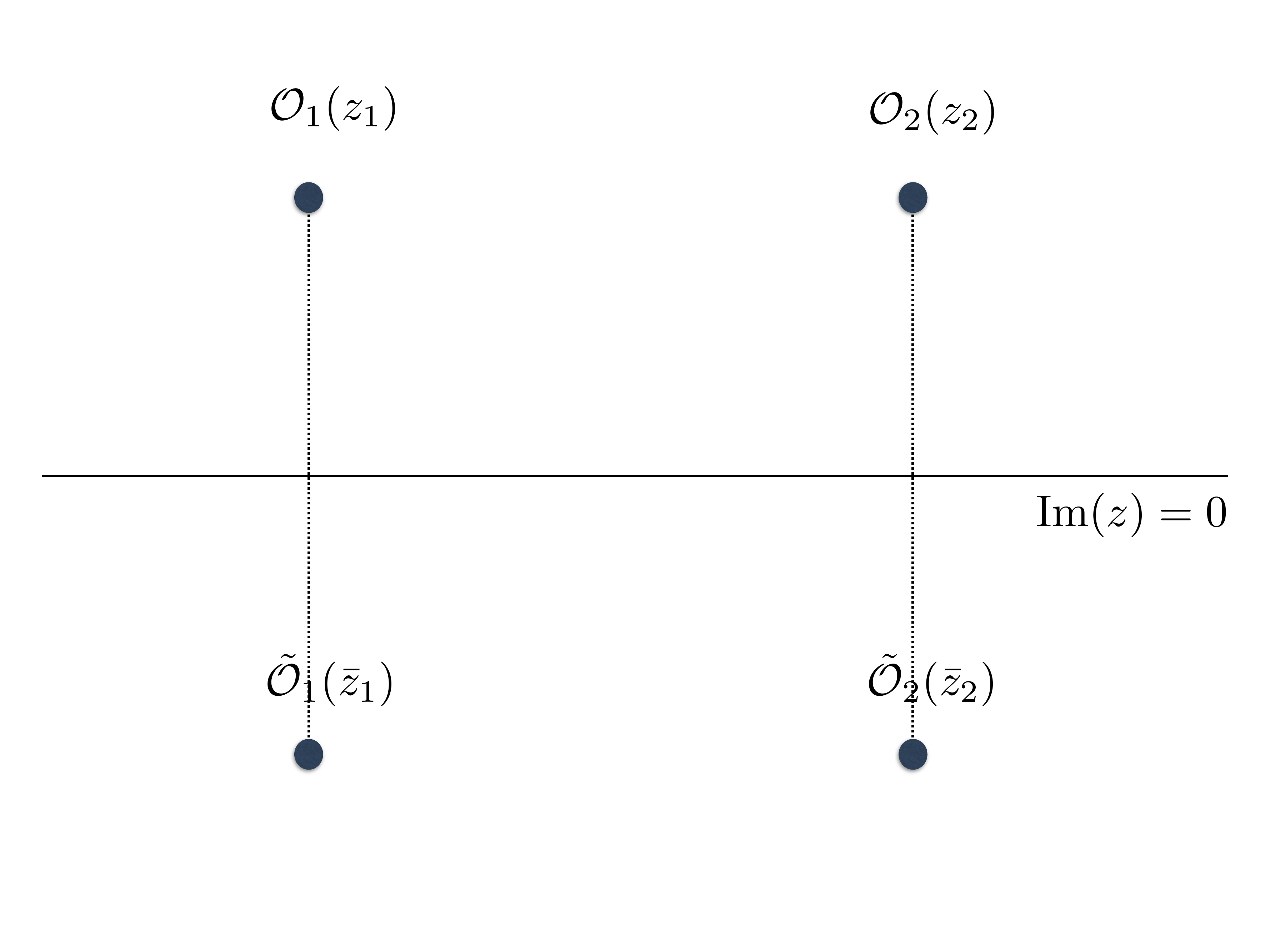}}
\caption{\small The simplest configuration in which a $2$-point correlator becomes a $4$-point correlator, in the presence of a boundary. The horizontal line, on which ${\rm Im}(z) =0$, denotes the boundary; $\cO_1$ and $\cO_2$ are two operators in the upper half plane (UHP). These operators have corresponding mirror images $\tilde{\cO_1}$ and $\tilde{\cO_2}$, located appropriately in the lower half plane. This particular $2$-point correlator, however, cannot be written as a $4$-point OTOC.} \label{2to4}
\end{center}
\end{figure}
\begin{figure}[ht!]
\begin{center}
{\includegraphics[width=0.8\textwidth]{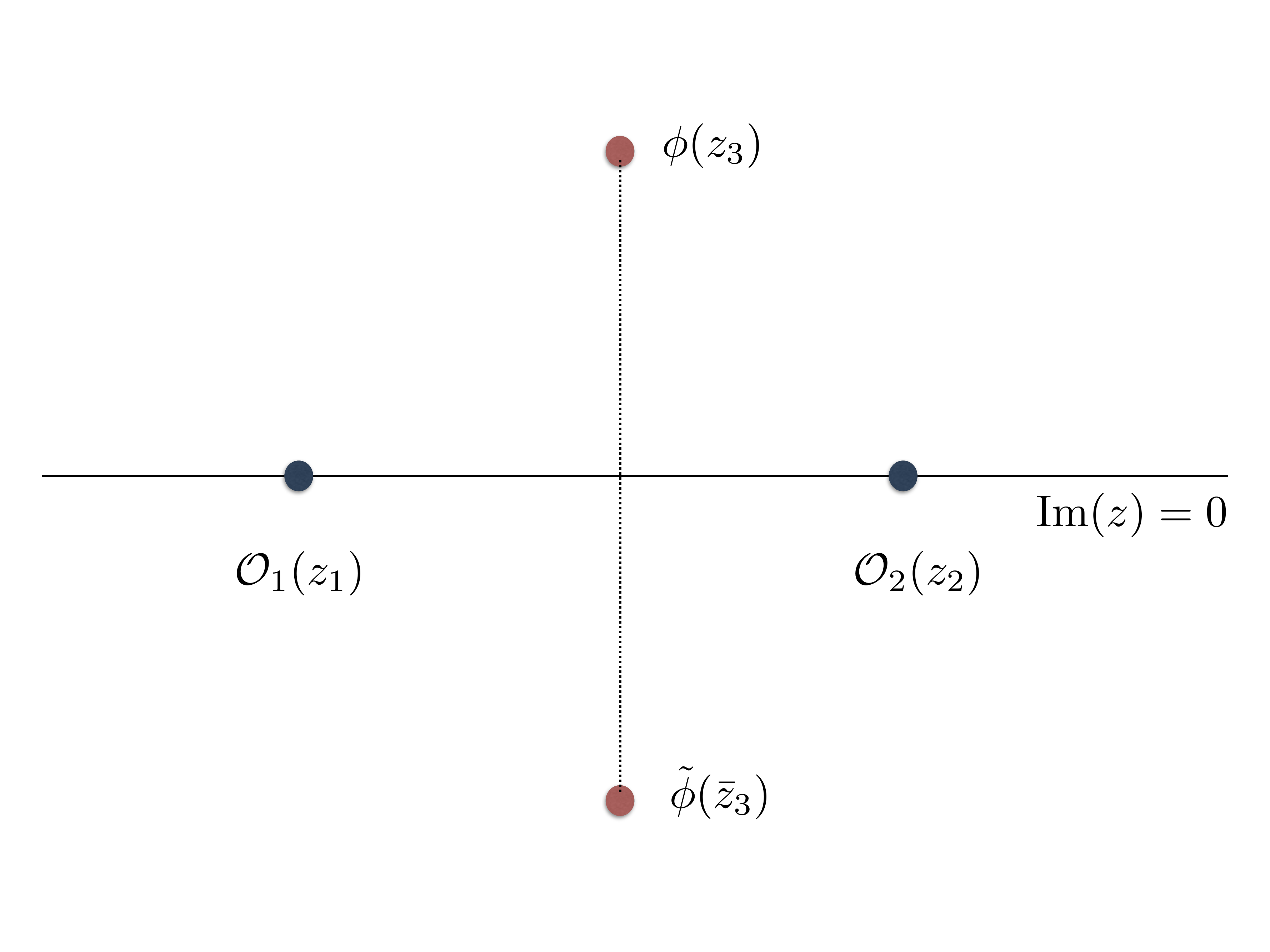}}
\caption{\small The non-trivial $3$-point correlator which becomes a $4$-point correlator, in the presence of a boundary. The horizontal line, as before, denotes the boundary; $\cO_1$ and $\cO_2$ are two operators inserted on the boundary, and $\phi$ is an operator inserted in the bulk. The latter one has a corresponding mirror image $\tilde{\phi}$, located appropriately in the lower half plane. This particular $3$-point correlator in the BCFT can be written as a non-trivial $4$-point OTOC in the plane. Here, $z_1$ and $z_2$ both lie on the ${\rm Im}(z) =0$ line.} \label{3to4}
\end{center}
\end{figure}
Thermal states are, however, special. This is due to the Kubo-Martin-Schwinger (KMS) condition. It can be easily shown\footnote{We have reviewed these arguments in an appendix later.} that the $3$-point thermal correlator, with an arbitrary time-ordering can be reduced to a TOC, by repeated use of the KMS condition. Thus, in general, it is only $4$-point function onwards when OTOCs become relevant. The presence of a non-trivial boundary, however, can violate the KMS condition and make the $3$-point OTOC non-trivial, by simply having non-trivial operators sitting at the boundary. It is therefore an interesting question whether such a non-trivial $3$-point OTOC can capture the same information as an otherwise standard $4$-point OTOC.\footnote{See {\it e.g.}~\cite{Gharibyan:2019sag}, in which a similar question has been addressed, from a different perspective.} We will address this question in this article, with an affirmative answer.

Motivated by this result, it is natural to wonder whether $3$-point and $2$-point correlators, and low point correlation functions in general, can have dynamical imprints of higher point correlators, and if so, to what extent and in what structure. This is a broad question, and we will touch upon this question from the perspective of the recently observed curious phenomenon of ``pole-skipping". The main observation is related to the analytic properties of retarded Green's functions, which is essentially a $2$-point correlator. It was pointed out in \cite{Grozdanov:2017ajz}, and elaborated further on  \cite{Blake:2018leo, Blake:2017ris, Grozdanov:2018kkt, Blake:2019otz}, that the retarded correlator, of Holographic systems, takes a special form and the corresponding Green's function has a pole and a zero intersecting at special points in the $(\omega, k)$-plane, which results in the disappearance of certain poles in the retarded Green's function. This phenomenon is termed as the ``pole-skipping". Subsequently, it was pointed out that the pole-skipping is related to the behaviour of the OTOCs of the corresponding system. 

Given the above observation, in this article, we also explore the pole-skipping phenomenon for a generic CFT, with and without the presence of a boundary. We explore varied possibilities, and the phenomenon seems generic and natural in this context, similar to the generic nature of this behaviour within hydrodynamic description\cite{Grozdanov:2019uhi}. Furthermore, we find that the pole skipping phenomenon in a certain class of $2$-point functions in a CFT with a boundary does indeed contain the formation of the maximal Lyapunov exponent, and therefore, when applicable, the information of the chaotic behaviour of the same.

This article is divided into the following parts: In the next section we begin with a review of the analytic structure of a $4$-point OTOC in a two-dimensional CFT. In section $3$, we discuss the analytic structure of a $3$-point correlator in a BCFT, and show how this can lead to a maximal Lyapunov exponent with the example of Virasoro blocks. Section $4$ is devoted to the study of the pole-skipping phenomenon for conserved currents in a two-dimensional CFT and in section $5$ we present an explicit example of the same phenomenon in the $2$-point correlator in a BCFT. Finally, we conclude in section $6$; we have relegated various technical details in five appendices.

\section{Analytic structure \& $4$-pt OTOC in 2D CFT}

Correlation functions in a Lorentzian QFT can be obtained from its Euclidean counterpart, via an analytic continuation in complex time. The Euclidean functions are single valued and symmetric functions of the coordinates of the operators. However, as a function of the complex time, the correlation function will have in general a rich analytic structure with branch-cuts along the light-cones of the inserted operators. The existence of these branch-cuts implies that the correlation functions are multivalued along the cut. Depending on how the branch cut is crossed, we get differently ordered (Time Ordered, Out of Time Ordered) correlation functions in the Lorentzian theory. Operationally, the way to do this is to start with the Euclidean correlators, with operators inserted at small but purely imaginary times $\tau_i = 0+ i\epsilon_i$. Then  continue the time coordinates to their Lorentzian values $\tau_i = t_i +i\epsilon_i$  by finally taking the $\epsilon_i\rightarrow 0$ limit. The ordering of the operators in the lorentzian correlator is related to the order in which the $\epsilon_i\rightarrow 0$, with $\langle\mathcal{O}(t_1)\mathcal{O}(t_2)...\mathcal{O}(t_n)\rangle $, corresponding to the ordering $\epsilon_1<\epsilon_2<\epsilon_3<....<\epsilon_n$ \cite{Osterwalder:1973dx},\cite{Haag:1992hx} \footnote{See also \cite{Luscher:1974ez} and \cite{Hartman:2015lfa} for its application in the context of conformal field theory.}. 

In\cite{Roberts:2014ifa}, the authors studied the analytic continuation in the thermal 4-pt correlation function in a 2D CFT. Holographic CFT's with a weakly coupled bulk gravity dual, are maximally chaotic. This was established originally from bulk computations in\cite{Shenker:2013pqa, Shenker:2013yza, Shenker:2014cwa, Maldacena:2015waa}. The goal of\cite{Roberts:2014ifa} was to see the butterfly effect in a direct field theoretic computation of the OTOCs.\footnote{See {\it e.g.}~\cite{Stanford:2015owe, Steinberg:2019uqb} a perturbative QFT computation of OTOC, and the corresponding Lyapunov exponent.} To this end, they study the behaviour of thermal expectation values of out of time ordered operators. In a 2D CFT, the thermal correlation functions (in a cylindrical geometry) are related by a conformal map to vaccum correlation functions on the plane. On the plane, the identity Virasoro conformal block contribution to the 4-pt function in a large $c$ limit, is expected to reproduce the bulk Einstein gravity computations, thus providing a field theoretic counterpart of the holographic computations done to study the behaviour of chaos in thermal systems.

The object of their study was the following thermal OTOC: $\langle W(x, 0)V(0, t)W(x, 0)V(0, t)\rangle_{\beta}$, where $<>_{\beta}$ denotes the thermal expectation value at temperature $T=\frac{2\pi}{\beta}$, while $W$ and $V$ are primary operators with conformal dimensions ($h_{W}$, $\bar{h}_{W}$) and ($h_{V}$, $\bar{h}_{V}$). As mentioned earlier, this is evaluated via an analytic continuation from the following Euclidean correlator $\langle W(x, 0+i\epsilon_1)W(x, 0+i\epsilon_2)V(0, 0+ i\epsilon_3)V(0, 0+ i\epsilon_4) \rangle_{\beta}$, and then continuing in the real time coordinate of the $V$ operator from $0\rightarrow t$. Finally to get to the OTOC, we take $\epsilon_{i}\rightarrow 0$ in the following order $\epsilon_{1}<\epsilon_{3}<\epsilon_{2}<\epsilon_4$. 

The (normalized) 4-pt function of a CFT on a plane, has the following form fixed by conformal invariance:
\begin{equation}\label{butterfly}
\frac{\langle W(z_1, \bar{z}_1)W(z_2,\bar{z}_2)V(z_3, \bar{z}_3)V(z_4, \bar{z}_4)\rangle}{\langle W(z_1,\bar{z}_1)W(z_,\bar{z}_2)\rangle\langle V(z_3,\bar{z}_3)V(z_4,\bar{z}_4)\rangle} \propto f(z,\bar{z}) \ ,
\end{equation} 
$z$ is the cross-ratio $z =\frac{z_{12}z_{34}}{z_{13}z_{24}}$. As mentioned previously, thermal correlation functions are related by a conformal map to vacuum correlation functions on the plane. The conformal map relating the coordinates $z_i$ on the plane to the (complex) coordinates ($\alpha_i$) on the cylinder is given by: $z_i = e^{\frac{2\pi\alpha_i}{\beta}}$, with $\alpha_i = x_i+t'_i +i\epsilon_i$. This map is valid at all values of the complex time. The analytic continuation is done in the time coordinate $(t')$ of the $V$ operator, while the $W$ operator is placed at $t'=0$. We are interested in the behaviour of the correlation function at large $t$. Analytic continuation in $t'$ is also an analytic continuation in the cross-ratio $z$. The final form of the 4-pt function and in particular the function of the cross ratio $f(z,\bar{z})$ would crucially depend on the path taken during the analytic continuation. 

The explicit form of $f(z,\bar{z})$ would of course be theory dependent. When expanded in the s-channel, this function gets contribution from the Virasoro conformal blocks associated to the various primary fields in the spectrum of the specific theory. However the identity block will always be exchanged in the s-channel of this four point function (as shown in Figure 1), and which in holographic CFT's in the large central charge limit, captures several features of the dual AdS$_3$ Gravity sector\cite{Hartman:2013mia},\cite{Fitzpatrick:2014vua},\cite{Fitzpatrick:2015zha}.

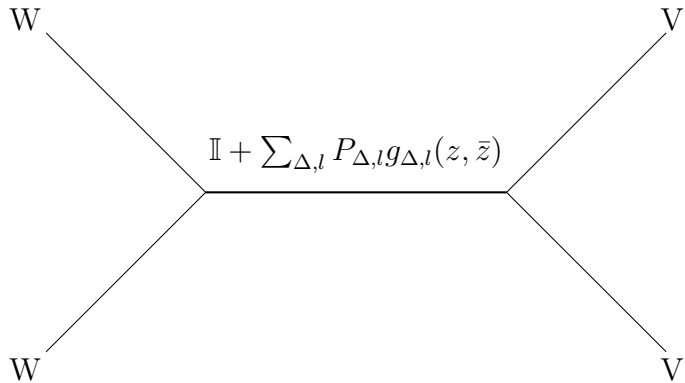
\begin{figure}
\begin{center}
\begin{tikzpicture}
\draw[thick] (-2.5,-5.5) --+(0:4);
\draw (-2.5,-5.5) --+(135:3);
\draw (-2.5,-5.5) --+(225:3);
\draw (1.5,-5.5) --+(45:3);
\draw (1.5,-5.5) --+(315:3);
\node (W) at (-4.9,-3.2) {W};
\node (W) at (-4.9,-7.85) {W};
\node (V) at (3.7,-3.2) {V};
\node (V) at (3.7,-7.85) {V};
\node at (-0.5,-5) {$\mathbb{I}+\sum_{\Delta,l}P_{\Delta,l}g_{\Delta,l}(z,\bar{z})$};
\end{tikzpicture}
\end{center}
\caption{$\mathbb{I}$ is the Virasoro identity block coming purely from the vacuum sector. Contributions from other primaries(and their descendants) belong to the sum over blocks $g_{\Delta,l}(z,\bar{z})$. }
\end{figure}

Thus any bulk gravity computation of OTOC's is expected to be reproduced from this identity block sector of the four-pt function. To this end, the authors of \cite{Roberts:2014ifa} study the analytic structure of the identity Virasoro conformal block ($\mathcal{F}_{0}(z)$) in the 2D CFT at large c. While the Virasoro conformal blocks can be computed as a series expansion \cite{Ferrara:1974ny, Zamolodchikov:1985ie}, there is no known closed form expression for the same, except in some limiting cases. In particular, in the large $c$ limit, with  $h_w/c$  fixed at a  small value and keeping $h_v$ fixed and large, it is known to have the following explicit form: $\mathcal{F}_{0}(z)\approx \Big(\frac{z}{1-(1-z)^{(1-12h_w/c)}} \Big)^{2h_v}$ \cite{Fitzpatrick:2014vua, Fitzpatrick:2015zha}.     \\

\noindent {\bf Analytic continuation in the identity block: }\\

Taking into account only the identity block sector, the function $f(z,\bar{z})$ takes the following form:
\begin{equation}
f(z,\bar{z}) \rightarrow \mathcal{F}_0(z)\mathcal{F}_{0}(\bar{z}) \nonumber \ .
\end{equation}

With this explicit form, we can now see how the analytic continuation happens in $z$ for the TOC and OTOC cases. As we analytically continue in time from $t'=0$ to $t'$ large, the corresponding path traced out in the $z$ plane is that of a closed loop from the origin of the $z$ plane (corresponding to the value at  $t'=0$) and back to the origin (corresponding to $t'$ large). The function $f$ has branch points at zero and one.   The difference in behaviour of the differently ordered correlation functions comes about in the way the analytically continued path crosses the corresponding branch cuts. The important question is whether the path taken during the analytic continuation cuts the x-axis of the complex $z$-plane along the branch cut or away from  it. This will lead to different functional dependence of the function $f(z,\bar{z})$ at large $t$.  Its easy to see that  at $t'\rightarrow x$, $z$ approaches the x-axis and cuts it at $z = \frac{\epsilon_{12}\epsilon_{34}}{\epsilon_{13}\epsilon_{24}}$. Precisely for the OTOC cases, say for instance in the case when $\epsilon_{1}<\epsilon_3<\epsilon_2<\epsilon_4$, we see that at this value of $(t'=x)$, the path crosses the branch cut along the positive $x$ axis of the $z$ plane, starting from $(z=1\rightarrow \infty)$ while for the TOC cases it does not. As a result, the functional dependence of the TOC's and the OTOC's at large $t$ are very different. 

In particular, we see that for the OTOC's  for which case the path crosses the branch cut, the factor $(1-z)$ in the expression for $\mathcal{F}_{0}(z)$, gets an extra phase factor of $e^{2\pi(1-12h_w/c)}$. In the case of the TOC's when the path does not cut the branch cut, one does not get the extra phase factor. The end result of this analysis is that at large $t$, the $f(z,\bar{z})$ has a non trivial dependence on $t$ for large $t$. On the other hand for the TOC's we have $f(z,\bar{z})\propto 1$. It is easy to see that the analytical continuation in the $\bar{z}$ plane does not cross the $\bar{z}=1$ branch point in either the TOC or the OTOC cases. Hence the final answer for $f$ is given by
\begin{equation}
f \approx \left( \frac{1}{1+ae^{\frac{2\pi}{\beta}(t-t_* -x)}} \right)^{2h_v} \ , \label{fex}
\end{equation}
where $a=\frac{24\pi ih_w}{\epsilon^*_{12}\epsilon_{34}}$ and $t_* =\frac{\beta}{2\pi}\log c$. In \cite{Roberts:2014ifa}, this was then exactly matched with the bulk computation of a two-pt function of a bulk operator corresponding to $V$ in a shock wave geometry created by the $W$ operator.

\section{Three pt functions in BCFT}

Consider a 2D Euclidean BCFT, at finite temperature. For simplicity, we chose a system which lives on a half line $\infty<x<0$. Then the 2D geometry is a cylinder, with the radius of the circle being $\beta/2\pi$. This system can be mapped to the upper half plane (UHP), by a conformal transformation. The map can be visualized as a two-step conformal transformation, mapping the cylinder to the unit disc and then followed by a map from the unit disc to the UHP.

\noindent {\bf Map from the cylinder to the unit disc:} 

Let $\alpha$ and $\bar{\alpha}$ be the complex coordinates on the cylinder and $\omega$ and $\bar{\omega}$ be the same on the unit disc, then in terms of these, the map from the cylinder to the  unit disc is given as follows:
\begin{equation}\label{map-1}
\omega = e^{\frac{2\pi}{\beta}\alpha} ; \quad \bar{\omega} = e^{\frac{2\pi}{\beta}\bar{\alpha}} \; \; \; |\omega | \leq 1 \ .
\end{equation}

\noindent {\bf Map from the unit disc to the UHP:}

The map from the unit disc to the UHP described by the coordinates ($z$,$\bar{z}$), is given by
\begin{equation}\label{map-2}
z= i\frac{1+\omega}{1-\omega}, \bar{z}= -i\frac{1+\bar{\omega}}{1-\bar{\omega}}, \; \; \; \textrm{Im}z\geq 0 \ .
\end{equation}

Combining them together, we get the full map from the cylinder to the UHP:
\begin{equation}\label{map}
z= i\frac{1+ e^{\frac{2\pi}{\beta}\alpha}}{1- e^{\frac{2\pi}{\beta}\alpha}},\; \; \bar{z}= -i\frac{1+e^{\frac{2\pi}{\beta}\bar{\alpha}}}{1-e^{\frac{2\pi}{\beta}\bar{\alpha}}}, \; \; \; \textrm{Im}z\geq 0 \ . 
\end{equation}
We will be interested in analytic continuations (in time) from the Euclidean to the Lorentzian cylinder. The above transformations would be valid throughout the complex time plane (t+i$\tau$), with $\alpha= x+ t+i\tau$, where $t$ is the Lorentzian time. 

Generally in a CFT, the three and two point functions have a very simple structure, which is completely determined upto a constant by conformal symmetry. Due to this fact, they do not show any interesting features under the analytic continuation in complex time from the Euclidean to Lorentzian time. However, in the BCFT, due to the reduced symmetry the two and three point functions have a non trivial and interesting structure. In particular, it is known that the conformal ward identities imply that the n-pt function in the BCFT, behaves as a 2n-pt function of a holomorphic bulk cft, under conformal transformations. 

It is with this motivation that we now look at the analytic structure of an OTOC  3-pt function, with two boundary operators and one bulk operator in BCFT.  While it is true that the 2-pt function in the BCFT, pictorially shown in figure \ref{2to4}, will have the structure of a 4-pt function in a holomorphic bulk CFT, a two-pt function cannot be a OTOC, so this is not of much use to us. Nevertheless we will discuss the analytic structure of these functions in the Appendix.


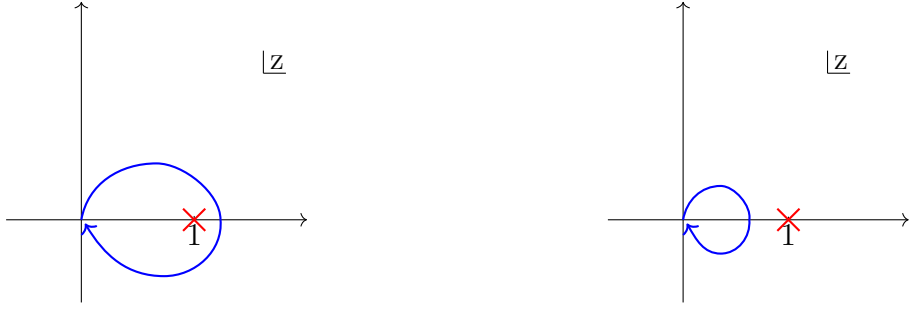
\begin{figure}
\begin{center}
\begin{tikzpicture}

\draw[->] (-3.5,0) --+(0:4);
\draw[->] (-2.5,-1.1) --+(90:4);
\draw (-1,-0.009)--+(90:0.06);
\node (1) at (-1,-0.2) {1};
\node (z) at (0.1,2.1) {z};
\draw (-0.08,1.95) --+(90:.3);
\draw (-0.08,1.95) --+(0:.3);
\draw[thick,blue,->] (-2.5,0) to[out=80,in=180,looseness=1] (-1.5,0.75) to[out=0,in=90,looseness=0.8] (-0.65,0) to[out=275,in=0,looseness=1] (-1.4,-0.75) 
to[out=180,in=-60,looseness=1] (-2.45,-0.05);
\draw[->] (4.5,0) --+(0:4);
\draw[->] (5.5,-1.1) --+(90:4);
\draw (6.9,-0.009)--+(90:0.05);
\node (1) at (6.9,-0.2) {1};
\node (z) at (7.6,2.1) {z};
\draw (7.42,1.95) --+(90:.3);
\draw (7.42,1.95) --+(0:.3);
\draw[thick,blue,->] (5.5,0) to[out=80,in=180,looseness=1] (6,0.45)
to[out=0,in=80,looseness=0.8] (6.38,0)
to[out=275,in=0,looseness=1] (6,-0.45)
to[out=180,in=-60,looseness=1] (5.55,-0.05);

\draw[thick] (-1,0) node[branch point,draw=red,thick] {};
\draw[thick] (6.9,0) node[branch point,draw=red,thick] {};


\end{tikzpicture}
\end{center}
\caption{On the right is shown the analytic continuation for the TOC cases while the diagram on th left shows the analytic continuation in the OTOC case.}
\end{figure}

\subsection{BCFT Bulk-boundary-boundary 3-pt function} 

One way to make a conformal 4-pt functions from BCFT lower point correlators is to consider bulk-boundary three point function. We take the bulk operator of dimension $h_{2}$ to be at ($-|x|, 0$) and the two boundary operators of dimension $h_{1}$ to be at ($0, t$). Using the 'doubling trick' and using equation(\ref{map}), we have the following four points on UHP and its mirror.
\begin{eqnarray}
&& z_{0} =i\frac{1+e^{-|x|+i\epsilon_0}}{1-e^{-|x|+i\epsilon_0}}, \; \; \bar{z}_{0} = -i\frac{1+e^{-|x|-i\epsilon_0}}{1-e^{-|x|-i\epsilon_0}} \nonumber \\
&& z_{1} = i\frac{1+e^{t +i\epsilon_1}}{1-e^{t+i\epsilon_1}}, \; \; z_2 = i\frac{1+e^{t+i\epsilon_2}}{1-e^{t +i\epsilon_2}} \ .
\end{eqnarray}

The cross ration $Z=\frac{(Z_0 -\bar{Z}_0)(Z_1 -Z_2)}{(Z_0 -Z_1)(\bar{Z}_0 -Z_2)}$ then turns out to be:
\begin{equation}
Z= i\epsilon_{12}\cdot e^{t}\cdot\frac{(1-e^{-2|x|})}{(e^{t}-e^{-|x|})(1-e^{t-|x|}e^{i\epsilon_{20}})} \ .
\end{equation}
Evaluating the cross ratio $Z$ for the three cases $t\rightarrow 0$, $t\rightarrow \infty$ and $t =|x|$, we get:
\begin{eqnarray}
&& Z_{t=0} = i\epsilon_{12} \Big(\frac{1+e^{-|x|}}{1-e^{-|x|}}\Big),\; Z_{t\rightarrow \infty} = -i\epsilon_{12}\cdot e^{|x|-t}(1-e^{-2|x|}) \ , \nonumber \\  
&& Z_{t=|x|} = \frac{\epsilon_{21}}{\epsilon_{20}} = 1 + \frac{\epsilon_{01}}{\epsilon_{20}} \ .
\end{eqnarray}
The cross ratio goes to zero from opposite directions in the $(t=0)$ and $t\rightarrow \infty$ limits. This is independent of the operator ordering.  However  at the point $(t =|x|)$, the behaviour of the cross ratio depends crucially on the ordering of the operators. 

Precisely for the OTO cases $\epsilon_2>\epsilon_0 > \epsilon_1$ and $ \epsilon_1>\epsilon_0 > \epsilon_2$, the $Z$ crosses the real axis after the ($Z=1$) branch point. In contrast, for the Time-Ordered cases it crosses the real axis before the ($Z=1$) branch point (See Figure 2). This behaviour is similar to the CFT four point function case as in \cite{Roberts:2014ifa}, even though in the BCFT it manifests itself at the three point function level.

\noindent {\bf large $c$ limit of a Vacuum Virasoro block} \\

We could again look at the large $c$ limit and take the same limit as in the CFT example, with $h_w/c$ fixed to a small value and $h_v$ fixed to a large value. We expect that in this limit again the only contribution would be from the vacuum global block. Therefore the analysis would be same as before. The contribution of $\mathcal{F}_{0}(Z)$ in the time ordered case at small $Z$ it gives 1. In the OTO case, the crossing of $Z=1$ branch point implies a non trivial contribution in the analytic continuation to the second sheet. Taking small $Z$, (\ref{fex}) reads as:
\begin{align}
\mathcal{F}_{0}(Z) \approx \left(\frac{1}{1-\frac{24\pi ih_{1}}{cZ}}\right)^{2h_{2}} \ .
\end{align}
The large time behaviour of this OTO suggests:
\begin{align}
\mathcal{F}_{0} \approx \left(\frac{1}{1+\frac{24\pi h_{1}}{\epsilon_{12}(1-e^{-2|x|})}e^{t-t_{*}-|x|}}\right)^{2h_{2}} \ .
\end{align}
Here $t_{*}$ is the scrambling time and in the conventional unit it takes the form as:
\begin{align}
t_{*} = \frac{\beta}{2\pi}\log c \ .
\end{align}

We conclude this part by noting that if the BCFT corresponds to the world-sheet of a open string theory in flat space, then the boundary would  correspond to the presence of a D-brane in the target space. The three point computation that we set up, would then correspond to  the three  point function of a bulk operator with two operators on the D-brane. The bulk operator could simply correspond to a graviton and the boundary operators are scalar, vector or spinor degrees of freedom. 

This is reminiscent of the framework studied {\it e.g.}~in \cite{Maldacena:2016upp, Murata:2017rbp, deBoer:2017xdk, Banerjee:2018twd, Banerjee:2018kwy}, in which the soft modes which are responsible for the maximal chaos comes from a graviton coupling of the D-brane degrees of freedom. In fact, it was shown explicitly in \cite{Banerjee:2018twd, Banerjee:2018kwy} that the degrees of freedom of a semi-classical string in a background AdS$_3$ couples to the gravity soft modes through a Schwarzian mode. The resulting coupling consists of two stringy degree of freedom and one gravity fluctuation.

\section{Pole Skipping in 2D CFT}

Let us review how pole skipping works in the two point function of a generic primary operator\footnote{Here, we will consider conserved currents only. For conserved currents, one has $\partial_\mu j^\mu = 0$. In the complex plane this takes the form: $\partial_{\bar{z}} j^{\bar{z}} + \partial_{z} j^{z} =0 $, which is readily solved by any holomorphic function $j$. Thus, holomorphicity implies conserved current. Here, we will mainly consider correlators of holomorphic primary operators.} in a two dimensional CFT. Even though in \cite{Grozdanov:2019uhi}, the pole-skipping phenomenon has already been explored for any CFT$_{2}$ 2-point correlator of scalars, we include our detailed analyses when the two operators correspond to conserved currents\footnote{Two point function with spin two and spin three conserved currents were already explored in \cite{Haehl:2018izb}}. This is done for the sake of completeness of the pole skipping literature in CFT$_{2}$, even though the existence of skipped poles in the two point functions of these operators is not in general related to chaos.  Consider a primary operator, $\cO$, with conformal dimension, $\Delta$ \footnote{i.e $h=\Delta$, $\bar{h}=0$}, the two point function in the complex-plane is given by
\begin{eqnarray}
\langle \cO(z_1) \cO(z_2) \rangle  = \frac{c}{\left( z_1 - z_2 \right) ^{2\Delta}} \ .
\end{eqnarray}
Now, perform the plane to cylinder map: $z = {\rm exp} (- i w)$, which corresponds to setting a temperature $\beta = 2 \pi$. Under the conformal transformation $z \to w$, a primary operator transforms as:
\begin{eqnarray}
\cO(w) = \left( \frac{\partial z(w)}{\partial w}\right)^\Delta \cO(z(w)) \ .
\end{eqnarray}
Thus, the two-point function will take the form:
\begin{eqnarray}
\langle \cO(w_1) \cO(w_2) \rangle  = \frac{(-1)^\Delta}{\left( 2 i \right)^{2\Delta} \sin^{2\Delta} \left( \frac{w_2-w_1}{2}\right)} \ .
\end{eqnarray}
We can write $w_2 - w_1 = \tau + i \sigma$, such that the correlator above is periodic under $\tau \to \tau + 2\pi$. The corresponding Euclidean correlator in the momentum space is now given by
\begin{eqnarray}
G_{\cO}^{\rm E} \left( \omega_{\rm E}, k\right)  = \langle \cO\left( \omega_{\rm E}, k \right) \cO \left( - \omega_{\rm E}, - k \right) \rangle  = \frac{(-1)^\Delta}{\left( 2 i \right)^{2\Delta}} \int d\tau d\sigma \frac{e^{- i \omega_{\rm E} \tau - i k \sigma}}{\sin^{2\Delta} \left( \frac{ \tau + i\sigma}{2}\right)}
\end{eqnarray}
Now, make the following change of integration variables:
\begin{eqnarray}
y = e^{i\tau} \ , \quad r = |\sigma| \ .
\end{eqnarray}
This yields:
\begin{eqnarray}
G_{\cO}^{\rm E} \left( \omega_{\rm E}, k\right)   = \frac{(-1)^\Delta}{i} \int_0^\infty dr \oint_{|y|=1} dy y^{\Delta - 1 - \omega_{\rm E}} \left[ \frac{e^{-(\Delta - ik) r}}{(y- e^{-r})^{2\Delta}} + \frac{e^{(\Delta - ik) r}}{(y- e^{r})^{2\Delta}} \right] \ . \label{GO}
\end{eqnarray}
The idea now, similar to \cite{Haehl:2018izb}, is to evaluate the Euclidean correlator above and analyze its' analytic properties. We have considered various possibilities, in details, in appendix B and here we summarize our observations, in the table below:
\begin{center}
\begin{tabular}{ |c|c|c|c| } 
 \hline
 Cases & Sub-cases & Skippable Poles  & Pole-skipping at \\
 \hline
$ \Delta = n+1$ , $n\in {\mathbb Z}_+$ & $\Delta - \omega_{\rm E} > 1$ & $\omega_{\rm E} = ik$ & $\omega_{\rm E} = \pm j$ , $j = 0, \ldots , n$ \\ 
$ \Delta = n+\frac{1}{2}$ , $n\in {\mathbb Z}_+$ & $\Delta - \omega_{\rm E} > 1$ & $\omega_{\rm E} = ik$ & $\omega_{\rm E} = \pm \left( j + \frac{1}{2} \right) $ , $j = 0, \ldots , n$ \\ 
$ 2\Delta \not= {\rm Integer}$, $m\in {\mathbb Z}_+$  & $\Delta - \omega_{\rm E} -1 = {\rm Integer}>0$ & $\omega_{\rm E} = ik$  & $\omega_{\rm E} = \frac{1}{2} \left(\Delta + ik - m -1 \right) $  \\ 
& $\Delta - \omega_{\rm E} -1 = {\rm Integer} < 0$ & $\omega_{\rm E} = ik$ and others & $\omega_{\rm E} = 2 + n - ik $ \\
& $\Delta - \omega_{\rm E} -1 \not= {\rm Integer} $ & $\omega_{\rm E} = ik$ and others & no pole-skipping \\
 \hline
\end{tabular}\label{pskiptable}
\end{center}

Let us offer a few comments regarding our observations. First, we have grouped the results in two categories: when $2\Delta$ is integer-valued and when it is not\footnote{Existence of non-local conserved currents with fractional spin (i.e with non integer $\Delta$ in our case) was observed in \cite{Fateev:1985mm}.}. We have not considered $\Delta <0$, which violates unitarity bound for CFTs.\footnote{Outside the unitarity bound, the pole-skipping phenomenon seems to persist still.} When $2\Delta$ is integer-valued, generic pole-skipping occurs for both integer and half-integer valued Matsubara frequencies. Furthermore, when $2\Delta$ is non-integer, pole-skipping is also observed at non-integer values of the frequency.

Typically, for an Euclidean system, non-integer values of $\omega_{\rm E}$ lacks regularity and, therefore, corresponds to systems which are not in thermal equilibrium with the bath temperature. Thus, our observations indicate that pole-skipping may persist arbitrarily far away from equilibrium. Since there is no natural {\it small parameter} associated with the non-integer values of $\omega_{\rm E}$, there is no natural way to associate a small departure from equilibrium for the corresponding physical mode. However, the physical meaning of the Euclidean correlator is unclear in this regime and one, instead, considers Wigner transformed spectral functions. Whether our observation can be made more precise and rigorous in the non-equilibrium framework, remains to be seen. We hope to come back to this issue in future.

Before concluding this section, let us note the following. Given a field $\phi$, defined on an Euclidean thermal circle, we can impose:
\begin{eqnarray}
\phi(\tau + \beta) = e^{i\theta} \phi(\tau) \ ,
\end{eqnarray}
where $\theta$ is a real parameter. For Bosonic fields, $\theta = 2\pi n$, $n \in {\mathbb Z}$ and for Fermionic fields $\theta = (2m + 1) \pi$, $m \in {\mathbb Z}$. The corresponding Matsubara modes are given by
\begin{eqnarray}
\omega_{\rm E} & = & \frac{2 \pi n}{\beta} \ , \quad {\rm Bosons} \\
& = & \frac{(2m + 1)}{\beta} \pi \ , \quad {\rm Fermions} \ , \\
& = &  \frac{2 \pi n}{\beta} + \frac{\theta}{\beta} \ , \quad {\rm generally} \ .
\end{eqnarray}
The last line above has a formal similarity to Matsubara modes of anyons, although we are working in a two-dimensional Euclidean framework. In the general case, clearly, the corresponding Matsubara modes can take non-integer values.

\section{Bulk-boundary two point function in BCFT}

\begin{figure}[ht!]
\begin{center}
{\includegraphics[width=0.8\textwidth]{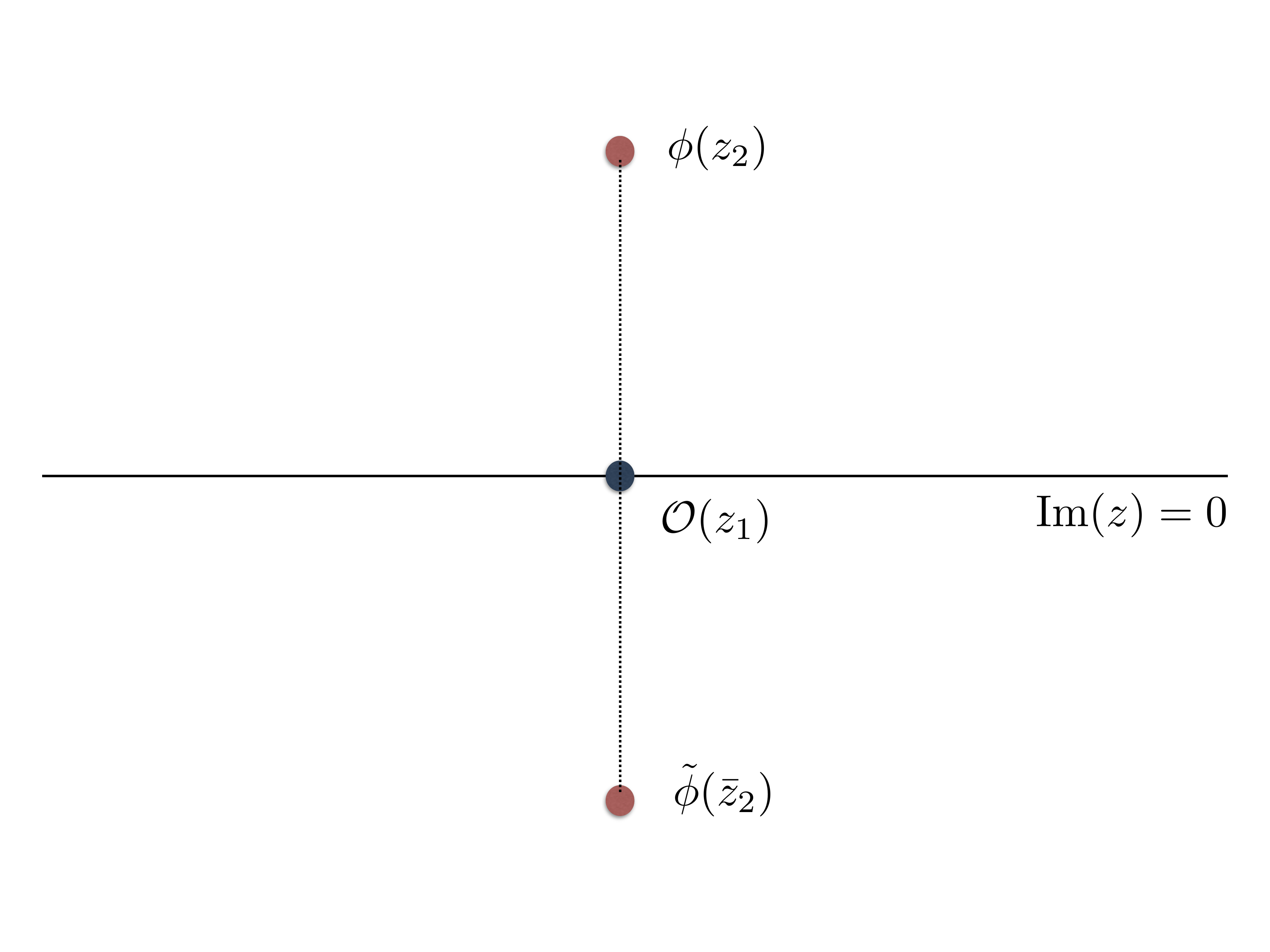}}
\caption{\small The simplest $2$-point correlator in a BCFT, which yields a non-trivial pole-skipping structure. The $2$-point correlator, with this structure, becomes a $3$-point correlator.} \label{2to3}
\end{center}
\end{figure}

In keeping with the observations made so far, let us explore whether the pole-skipping is observed in {\it low point} correlators in a BCFT. It is straightforward to check that an one-point function in the BCFT, which becomes a $2$-point correlator in the plane, does not reveal any pole-skipping behaviour. Therefore, let us consider the next non-trivial configuration shown in figure~\ref{2to3}.

This corresponds to a bulk-boundary two point function of bulk scalar field $\phi$ with dimension $\Delta_{1}$ and boundary field $\mathcal{O}$ with dimension $\Delta_{2}$ on UHP. Also let us assume in the Euclidean cylindrical coordinates, $\phi$ is located at $(-|x|,0)$ and $\mathcal{O}$ is located at $(0,\tau)$. Using method of images, this two point function could be evaluated by considering holomorphic part of three point function on the whole complex plane as follows:
\begin{align}
\braket {\phi(-|x|,0) \mathcal{O}(0,i\tau)} \approx \braket {\mathcal{O}(z_{0}) \mathcal{O}(\bar{z}_{0}) \mathcal{O}(z_{1})} \ .
\end{align}
Now, given the $3$-point correlator in the plane, we can compute the Fourier transform. We have presented the details of this calculation in appendix C. Here, let us summarize the observations.

For convenience, let us fix $\Delta_2=2$, which corresponds to the stress-tensor. Thus, the $2$-point BCFT correlator is between a bulk operator and a boundary stress-tensor insertion. In this case, as detailed in appendix C, pole skipping is observed on integer-valued positive Matsubara frequencies, irrespective of $\Delta_1$. Similarly, one can also check that for $\Delta_2=1$, there is no pole-skipping phenomenon. Thus, for the BCFT $2$-point correlator, one indeed observes a pole-skipping at Matsubara frequencies (the lowest of which is related to the maximal chaos), provided the operator at the boundary is chosen judiciously.

\section{Conclusions}

In this article, we have initiated studying the imprint of boundaries in the dynamics of correlation functions. We have presented a specific configuration of a $3$-point function in a BCFT which captures the essential dynamical content of an otherwise $4$-point correlators, including OTOC. Furthermore, we have presented evidence that, the recently observed phenomenon of pole-skipping in $2$ point correlator, in the presence of a boundary, is a generic feature. Our analyses here applies in the context of Holography, as well as within a purely CFT context. This certainly hints towards a potentially rich structure which is yet to be uncovered.

Naively, taking the observations made in this article in the most general context, it appears possible that non-trivial dynamical information of higher point correlation functions may, in certain intertwined manner, be already captured within low-point correlators. This statement, clearly, does not define what {\it higher point} or a {\it lower point} correlation function is; nor does it specify how the information is, in general, encoded. Our observed structure is mostly encoded in analytic properties of the lower point correlator, so it's possible that information content of higher point OTOC is buried in more and more complicated analytic continuation of a low point correlation function. This statement, at present, is completely speculative. For example, to the best of our knowledge, we are not aware of an LSZ-type reduction involving OTOCs in a QFT, which could be related to a standard LSZ-reduction {\it via} analytic continuation. It will be extremely interesting to explore some of these aspects in future.

A simple way in which a classical system can interpolate between being integrable to completely ergodic is by changing the boundary condition, appropriately. For example, a free particle without any boundaries is integrable; with a rhombic boundary condition it looses integrability but does not yet become chaotic; and with a billiard stadium boundary condition, it becomes chaotic. In terms of classical trajectories, this behaviour is intuitive: a free particle is completely solvable, a rhombic boundary condition effectively introduces a branch cut in the plane which destroys the integrability, and a billiard stadium induces defocussing of trajectories which can give rise to exponentially diverging nearby trajectories. In the semi-classical limit, whether an analogous statement can be made in terms of correlation functions, will be an extremely interesting issue to explore further.

Finally, note that, the phenomenon of pole-skipping is {\it a priori} unrelated to the ergodic nature of the system. As has been noticed earlier in \cite{Natsuume:2019sfp}, and as we have also discussed in this article, pole-skipping at Matsubara frequencies appear to be a generic and kinematic feature, irrespective of the CFT. Clearly, not all such systems are maximally chaotic; but for the ones which are, the lowest Matsubara frequency is related to the maximal Lyapunov exponent, {\it via} an analytic continuation. As an intriguing observation, it remains to be seen whether such a pole-skipping effect persists in a wider class of QFTs. We hope to address some of these issues in near future.

\section{Acknowledgements}

We would like to thank Gautam Mandal, Swarnendu Sarkar, Ritam Sinha, Shashi Srivastava for various communications and conversations. This work is supported by the Department of Atomic Energy. SD would like to thank The Abdus Salam International Center for Theoretical
Physics, Italy (Spring School on Super String Theory and Related Topics) for their warm hospitality during which, part of this work was completed. The work of SD was supported by a senior research fellowship(SRF) from CSIR.

\newpage
\appendix

\section{$2$-pt Function of a BCFT with Bulk Operators: Analytic Continuation}

Let us consider two bulk fields at the points $(x_{1},t_{1})$ and $(x_{2},t_{2})$. The map (\ref{map}), relates those points to UHP as following.
\begin{equation}
z_a =i\frac{1+e^{(x_{a}+t_{a}+i\epsilon_{a})}}{1- e^{(x_{a}+t_{a}+i\epsilon_{a})}}, \; \; \bar{z}_{a} =-i\frac{1+e^{(x_{a}-t_{a}-i\epsilon_{a})}}{1- e^{(x_{a}-t_{a}-i\epsilon_{a})}}, \; \; (a=1,2) \ .
\end{equation}
It follows from the 'doubling trick' that this 2-pt function transforms as a four point function in the full CFT, where the four fields are located at $z_{1},z_{2},\bar{z}_{1},\bar{z}_{2}$. This four point function is a function of cross ratio $Z =\frac{z_{1\bar{1}}z_{{2}\bar{2}}}{z_{12}z_{\bar{1}\bar{2}}}$, which in $x,t$ coordinates is given by:
\begin{eqnarray}
Z = && -\frac{(1-e^{2 x_1})(1-e^{2x_2})}{(e^{(x_{1}+t_{1}+i\epsilon_{1})}-e^{(x_{2}+t_{2}+i\epsilon_{2})})(e^{(x_{1}-t_{1}-i\epsilon_{1})}-e^{(x_{2}-t_{2}-i\epsilon_{2})})} \nonumber \\
=&& \frac{1}{A[B\cosh(t-i\epsilon)-1]} \ .
\end{eqnarray}
Where $A= \frac{e^{-(x_1+x_2)}\cosh(x_{12})}{2\sinh(x_1)\sinh(x_2)}$ and $B=\frac{1}{\cosh(x_{12})}$ and $t= t_1 -t_2$ and $\epsilon = \epsilon_1 -\epsilon_2$. We observe the following behaviour of cross ratio by taking different limits in $t$.

\begin{itemize}
\item For $t=0$, $Z= \frac{1}{A(B\cos(\epsilon)-1)}$. This is always a negative real number for any $A$ and $B$. 
\item As $t\rightarrow x_{12}$, $Z\rightarrow -\infty$. For $t > x_{12}$, $Z$ is positive. 
\item For $t\gg 1$, $Z= \frac{1}{AB}e^{-t}e^{i\epsilon}$.   Depending on the sign of $\epsilon$, which in turn denotes the ordering of the operators, $Z$ approaches $0$ from the first sheet or the second sheet.  
\end{itemize}
Therefore, in this case the trajectory of $Z$ does not cross branch point at $Z=1$. This is consistent with the fact that only for OTO's we expect a non trivial behaviour for the correlation functions at large time

\section{Pole-skipping in CFT: Generic Analyses}

\subsection{Poles}

Within the range of $r$, the second term above will not yield any residue when the $y$ integration is performed. The residue, therefore will come only from the first term inside the bracket. On the other hand, there may be a pole at $y=0$, provided 
\begin{eqnarray}
\Delta - \omega_{\rm E} -1 < 0 \quad {\rm and} \quad \Delta - \omega_{\rm E} -1 \in {\mathbb Z}_{-} \ .
\end{eqnarray}

Let us consider these two cases separately:

(i) Consider $\Delta > \omega_{\rm E} + 1$ and $\Delta \in {\mathbb Z}_+$. In this case, the only pole comes from the $(y - e^{-r})^{2\Delta}$ term inside the bracket and the corresponding integral is evaluated to be:
\begin{eqnarray}
&& G_{\cO}^{\rm E} \left( \omega_{\rm E}, k\right)   = \frac{(-1)^\Delta}{i} \int_0^\infty dr e^{(\omega_{\rm E} + ik) r} \Gamma\left( \Delta, \omega_{\rm E} \right) \ , \\
&& \Gamma\left( \Delta, \omega_{\rm E} \right) = \left( 2\pi i\right) \frac{\left( \Delta -\omega_{\rm E} - 1 \right)  \left( \Delta -\omega_{\rm E} - 2 \right)  \left( \Delta -\omega_{\rm E} - 3 \right) \ldots  \left(1- \Delta -\omega_{\rm E} \right) }{\left( 2\Delta - 1\right)!}\ . 
\end{eqnarray}
The $r$-integral can now be performed to yield:
\begin{eqnarray}
G_{\cO}^{\rm E} \left( \omega_{\rm E}, k\right) = \frac{(-1)^\Delta}{i} \left( - \frac{1}{\omega_{\rm E} + i k }\right) \Gamma\left( \Delta, \omega_{\rm E} \right) \ , \quad {\rm Re}\left( \omega_{\rm E}\right) < {\rm Im} \left( k \right) \ .
\end{eqnarray}
It is straightforward to check that, the pole along the line $\omega_{\rm E} + i k = 0$, in the correlator above, is skipped for $\Gamma (\Delta, \omega_{\rm E}) =0$. This yields the following modes, on which pole-skipping takes place:
\begin{eqnarray}
\omega_{\rm E} = \pm j \ , j = 0, \ldots, n \ , \quad \Delta = n+1 \ , \quad n \in {\mathbb Z}_+ \cup \{0\}\ .
\end{eqnarray}

Incidentally, for $\Delta \in \frac{1}{2}{\mathbb Z}_+$, the pole coming from $(y - e^{-r})^{2\Delta}$ term yields a non-trivial residue. In this case, one can repeat the exercise above and obtain the pole-skipping at:
\begin{eqnarray}
&& \omega_{\rm E} = \pm \left( j + \frac{1}{2} \right)  \ , \quad j = 0, \ldots, n \ , \\
&& \Delta = n+\frac{1}{2} \ , \quad n \in {\mathbb Z}_+ \cup \{0\}\ .
\end{eqnarray}

(ii) Consider now, $\Delta - \omega_{\rm E} - 1 = - p$ and $p \in {\mathbb Z}_+$. In this case, the residue also seems to exhibit the pole-skipping, only it's a bit subtle. We can analyze this case carefully. Towards that, let us write
\begin{eqnarray}
&& G_{\cO}^{\rm E} \left( \omega_{\rm E}, k\right)   = \frac{(-1)^\Delta}{i} \int_0^\infty dr I\left[ \Delta, \omega_{\rm E} \right] \ , \\
&& I\left[ \Delta, \omega_{\rm E} \right] = \left( 2\pi i \right) \left( {\rm Res} (y=0) + {\rm Res} (y=e^{-r}) \right) \ .
\end{eqnarray}
Here, $ I\left[ \Delta, \omega_{\rm E} \right]$ needs to be evaluated using the residue theorem at $y=0$ and $y = e^{-r}$. Now
\begin{eqnarray}
&& {\rm Res} (y=0) = \frac{(-1)^{p-2\omega_{\rm E}-1}}{(p-1)!} e^{-{\rm sgn}(\omega_{\rm E}) (\omega_{\rm E} + i k )r} \kappa(p, \omega_{\rm E})  \ , \\
&& \kappa(p, \omega_{\rm E}) = 2(p-\omega_{\rm E}-1) (2p-\omega_{\rm E}-3) \ldots (p-2\omega_{\rm E}) \ .
\end{eqnarray}
As we have evaluated earlier,
\begin{eqnarray}
&& {\rm Res} (y=e^{-r}) = \frac{1}{(2\omega_{\rm E}-2p)!} e^{-{\rm sgn}(\omega_{\rm E}) (\omega_{\rm E} + i k )r} \Gamma(p, \omega_{\rm E})  \ , \\
&& \Gamma(p, \omega_{\rm E}) = (-p)(-p-1) \ldots (p-2\omega_{\rm E}) \ .
\end{eqnarray}
Altogether, one obtains pole-skipping phenomena at discrete Matsubara frequencies, as observed in case (i) above.

Let us work out a specific example. Take $\Delta = 2$. In case (i), we get
\begin{eqnarray}
\Gamma(\omega_{\rm E}) = - \left( 2 \pi i \right) \frac{1}{3!} \omega_{\rm E} \left( \omega_{\rm E}^2 -1 \right) \ ,
\end{eqnarray}
and in case (ii) we get
\begin{eqnarray}
I = \left( 2 \pi i \right) \left[ - \frac{1}{3!}  \omega_{\rm E} \left( \omega_{\rm E}^2 -1 \right)  - \frac{1}{3!}  \omega_{\rm E} \left( \omega_{\rm E}^2 -1 \right) \right] \ .
\end{eqnarray}
Hence the pole-skipping takes place at $\omega_{\rm E} = 0 , \pm 1$. This matches with the result of \cite{Haehl:2018izb}.

Note that, there is no information of a large central charge limit in the above analyses, so it clearly holds for Holographic theories. However, given a CFT, with an upper bound on the spectrum, $\Delta_{\rm max}$, the pole-skipping is observed upto a maximum value of the Matsubara frequency. For holographic theories, this, along with the observation made in \cite{Blake:2019otz}, suggests that no such $\Delta_{\rm max}$ can exist.

\subsection{Branch-cuts}

Let us now consider non-integer values of $2\Delta$. In this case, there is no pole, but a branch cut coming from $(y - e^{-r})^{-2\Delta}$ in the denominator, as well as the $y^{\Delta - 1 -\omega_{\rm E}}$ from the numerator. We can choose representative values {\it e.g.}~$2\Delta = 1/2$. For now, let us assume $\omega_{\rm E} \in {\mathbb R}$.\footnote{Note that, for the regular Euclidean theory, $\omega_{\rm E} \in {\mathbb Z}$ or $\omega_{\rm E} \in \frac{1}{2}{\mathbb Z}$, because of the Euclidean periodicity. Relaxing this condition implies a loss of Euclidean periodicity and therefore corresponds to a generic non-equilibrium, non-thermal scenario.} The integrand in this case takes the form:
\begin{eqnarray}
\frac{1}{y^{3/4 + \omega_{\rm E}}} \frac{1}{\left( y - y_0 \right)^{1/2}} \ , \quad y_0 = e^{-r} \ .
\end{eqnarray}
The second factor will always have a branch-cut, running from $y_0$ to $\infty$, along the real axis.\footnote{This is a choice that we are allowed to make, without the loss of any generality. Note that, this feature is always true for any non-integer value of $2\Delta$.} The first factor has various options: if $\Delta - 1- \omega_{\rm E} \in {\mathbb Z_+}\cup \{0\}$, there is no branch cut, or pole; if $\Delta - 1- \omega_{\rm E} \in {\mathbb Z_-}$, there is a pole, but no branch-cut and if $\Delta - 1- \omega_{\rm E} \in {\mathbb R} \setminus {\mathbb Z}$, there is no pole, but a branch cut. Finally, there is another possibility: when $\Delta - 1 - \omega_{\rm E} \in {\mathbb R} \setminus {\mathbb Z}$, but $2\Delta \in {\mathbb Z}_+$.\footnote{We are excluding negative values of $\Delta$, which correspond to non-unitary theories.} Let us consider these cases separately.

(i) $\Delta - 1- \omega_{\rm E} \in {\mathbb Z_+} \cup \{0\}$: Let us write $\Delta - 1- \omega_{\rm E} = p$, where $p \ge 0$, and we will assume $\Delta =1/2$. The contour integral along the circle $|y|=1$ in (\ref{GO}) receives no contribution from the second term in the bracket in (\ref{GO}). Now, the branch-cut coming from $\left( y - y_0 \right)^{2\Delta}$ runs from $y_0$ to $\infty$, and therefore crosses the contour at $|y|=1$. The integral is ill-defined at that point because of the branch-cut and therefore we replace the integral by its principal value. 

We can pick a contour, as shown below. 
\begin{center}
\begin{tikzpicture}
  \tkzInit[xmin=-5,ymin=-5,xmax=5,ymax=5]
  \tkzDrawXY[noticks]
   \tkzDefPoint(0,0){O1}
\tkzDefPoint(1,0){O2}

   \tkzDefPoint(1.5,.2){B} \tkzDefPoint(1.5,-.2){C}
   \tkzDefPoint(4,.2){D} \tkzDefPoint(4,-.2){E} 
    
 \begin{scope}[decoration={markings,
      mark=at position .20 with {\arrow[scale=1]{>}};}]
  \tkzDrawArc[line width=1pt](O2,B)(C)
 \end{scope} 
  
  \begin{scope}[decoration={markings,
      mark=at position .5 with {\arrow[scale=1]{>}};}]
    \tkzDrawSegments[postaction={decorate},line width=1pt](B,D E,C)
  \end{scope}
  
  \begin{scope}[decoration={markings,
     mark=at position .20 with {\arrow[scale=1.5]{>}},
     mark=at position .80 with {\arrow[scale=1.5]{>}};
     }]
    \tkzDrawArc[postaction={decorate},line width=1pt](O1,D)(E)
  \end{scope}
  
  \draw[thick] (1,0) node[branch point,draw=red,thick] {};
  \draw[thick,red,branch cut] (1,0) to (0:5);
  
\node at (0.4,0.53) {$\gamma_{\varepsilon}$};
\node at (-3.4,2.8) {$\gamma_{R}$};
\node at (2.9,0.5) {$l_1$};
\node at (2.7,-0.5) {$l_2$};


\end{tikzpicture}
\end{center}
Thus, the contour integral, using the contour above, can now be written as:
\begin{eqnarray}
&& \I = \oint_{|y| =1} dy y^p \left[ \frac{y_0^{\Delta - ik }}{(y - y_0)^{2\Delta}} + \frac{y_0^{-\Delta + ik }}{(y - y_0^{-1})^{2\Delta}} \right] \ . \\
&& \I = \int_{\gamma_R + \gamma_\varepsilon} + \int_{l_1 + l_2} = 0 \ , \quad  {\rm P} (\I) = \int_{\gamma_R} \ , \label{pint}
\end{eqnarray}
where ${\rm P}(\I)$ denotes the principal value of the integral. On the segment $l_1+l_2$, the integral is evaluated as:
\begin{eqnarray}
&& y = y_0 + x e^{i \theta} \ , \\
&& l_1 : \theta = 2\pi \ , \quad \int_{l_1} = \int_{1-y_0}^0 dx (y_0+x)^p \frac{y_0^{\Delta - i k}}{x^{2\Delta} e^{4\pi i \Delta}}  \ , \\
&& l_2 : \theta = 0 \ , \quad \int_{l_2} = \int_0^{1-y_0} dx (y_0+x)^p \frac{y_0^{\Delta - i k}}{x^{2\Delta} }  \ , \\
&& \implies \quad \int_{l_1+l_2} = \left( 1 - e^{-4\pi i \Delta} \right) \int_0^{1-y_0} dx (x + y_0)^p \, \frac{y_0^{\Delta - ik}}{x^{2\Delta}} \ .
\end{eqnarray}
Also, the small circle around the branch point at $y_0$ yields:
\begin{eqnarray}
\int_{\gamma_\varepsilon} = \lim_{\epsilon\to 0} \int_{2\pi}^0 d\left( \epsilon e^{i\theta} \right) \left( \epsilon e^{i\theta} + y_0\right)^p \, \frac{y_0^{\Delta - ik}}{\epsilon^{2\Delta} e^{2i \Delta \theta}}  & = & \lim_{\epsilon\to  0} \epsilon^{1-2\Delta} (\ldots) \nonumber\\
& = & 0 \ , \quad {\rm if} \quad 1 > 2\Delta  \nonumber\\
& = & \infty \ , \quad {\rm if} \quad 1 < 2\Delta 
\end{eqnarray}
Thus, using (\ref{pint}), we get:
\begin{eqnarray}
P(\I) & = & - \left( 1 - e^{-4\pi i \Delta} \right) \int_0^{1-y_0} dx (x + y_0)^p \, \frac{y_0^{\Delta - ik}}{x^{2\Delta}} \ , \quad 1 > 2\Delta \ , \\
& = &  \left( e^{-4\pi i \Delta} -1 \right)\frac{\Gamma (1-2 \Delta )}{\Gamma (2-2 \Delta )} (1-y_0)^{1-2 \Delta } \, _2F_1\left(-p,1-2 \Delta ;2-2 \Delta;\frac{y_0-1}{y_0}\right) y_0^{\Delta -i k+p} \nonumber\\
\end{eqnarray}
Now, one can perform the $r$ integral in (\ref{GO}). This yields:
\begin{eqnarray}
G_{\cO}^{\rm E} \left( \omega_{\rm E}, k\right) = \alpha \left( e^{- 4\pi i \omega_{\rm E}} -1 \right) \frac{\Gamma\left( p+1 -i k - \omega_{\rm E}\right) \Gamma\left( -2p - 1 - 2 \omega_{\rm E}\right)}{\Gamma\left( -p -i k + \omega_{\rm E}\right)\left(\omega_{\rm E} - ik \right) } \ , \quad p+1 + {\rm Im}(k) > {\rm Re}(\omega_{\rm E}) \ .
\end{eqnarray}
Here $\alpha$ is some numerical factor which is not important for us. The factor $\left( e^{- 4\pi i \omega_{\rm E}} -1 \right)$ is purely kinematic and independent of $p$. The poles are located at $\omega_{\rm E} - i k = 0$ as well as at the poles of the Gamma functions in the numerator: $p+1 -i k - \omega_{\rm E} = - m$ and $-2p - 1 - 2 \omega_{\rm E} = - m$\footnote{These correspond to poles at $\omega_{\rm E} = \frac{1}{2} (\Delta -i k+m)$ and $\Delta = \frac{m+1}{2}$, respectively.}, where $m$ denotes a positive definite integer. Such poles can be skipped for: 
\begin{eqnarray}
e^{- 4\pi i \omega_{\rm E}} -1 = 0 \ , \quad \implies \quad \omega_{\rm E} \in  \mathbb{Z} \cup \frac{1}{2} \mathbb{Z} \ .
\end{eqnarray}
These are precisely the Matsubara frequencies, for bosonic and fermionic degrees of freedom. Note, however, that the relation 
\begin{eqnarray}
\Delta - 1 - \omega_{\rm E} = p  \label{omegap}
\end{eqnarray}
has no solution for integer or half integer values of $\omega_{\rm E}$ and $p$, when $2\Delta$ is not an integer. Thus, constrained by (\ref{omegap}), there will be no pole skipping in this case, for the Euclidean theory. The only other source of a zero in the numerator can come from the pole of the Gamma function in the denominator, when $-p -i k + \omega_{\rm E} = - m$, which yields: 
\begin{eqnarray}
\omega_{\rm E} = \frac{1}{2} (\Delta +i k-m-1) \ .
\end{eqnarray}
The above equation certainly has solutions for integer as well as non-integer values of $\omega_{\rm E}$. The location of the pole, which can be skipped by the above condition, is given by
\begin{eqnarray}
\omega_{\rm E} = i k \ .
\end{eqnarray}
The other poles cannot be skipped.

(ii) $\Delta - 1- \omega_{\rm E} \in {\mathbb Z_-}$: The corresponding contour is given below: 
\begin{center}
\begin{tikzpicture}
  \tkzInit[xmin=-5,ymin=-5,xmax=5,ymax=5]

  \tkzDrawXY[noticks]
   \tkzDefPoint(0,0){O1}
\tkzDefPoint(1,0){O2}

   \tkzDefPoint(1.5,.2){B} \tkzDefPoint(1.5,-.2){C}
   \tkzDefPoint(4,.2){D} \tkzDefPoint(4,-.2){E} 
    
 \begin{scope}[decoration={markings,
      mark=at position .20 with {\arrow[scale=1]{>}};}]
  \tkzDrawArc[line width=1pt](O2,B)(C)
 \end{scope} 
  
  \begin{scope}[decoration={markings,
      mark=at position .5 with {\arrow[scale=1]{>}};}]
    \tkzDrawSegments[postaction={decorate},line width=1pt](B,D E,C)
  \end{scope}
  
  \begin{scope}[decoration={markings,
     mark=at position .20 with {\arrow[scale=1.5]{>}},
     mark=at position .80 with {\arrow[scale=1.5]{>}};
     }]
    \tkzDrawArc[postaction={decorate},line width=1pt](O1,D)(E)
  \end{scope}
  
  \draw[thick] (1,0) node[branch point,draw=red,thick] {};
  \draw[thick,red,branch cut] (1,0) to (0:5);
  
\node at (0.4,0.53) {$\gamma_{\varepsilon}$};
\node at (-3.4,2.8) {$\gamma_{R}$};
\node at (2.9,0.5) {$l_1$};
\node at (2.7,-0.5) {$l_2$};

\filldraw[draw=red,fill=red] (0,0.5\ht0) circle[radius=.25em];


\end{tikzpicture}
\end{center}
The integral needs to be replaced by its principal value, and is evaluated to be:
\begin{eqnarray}
P(\I) = 2 \pi i \left[ {\rm Res}(y=0)\right] - \int_{l_1 + l_2} - \int_{\gamma_\varepsilon} \ .
\end{eqnarray}
By similar arguments as above, we get:
\begin{eqnarray}
\int_{\gamma_\varepsilon} \sim \lim_{\epsilon\to 0}\epsilon^{1-2\Delta} \to 0 \ , \quad {\rm if} \quad 1 - 2\Delta > 0 \ .
\end{eqnarray}
It is now straightforward to check that:
\begin{eqnarray}
 - \int_{l_1 + l_2} = \left( e^{-4\pi i \Delta} -1 \right) \int_0^{1-y_0} dx (x + y_0)^p \, \frac{y_0^{\Delta - ik}}{x^{2\Delta}} \ , \quad 1 > 2\Delta \ ,
\end{eqnarray}
and 
\begin{eqnarray}
{\rm Res}(y=0) = \frac{1}{(p-1)!} \left[ (-2 \Delta) (-2 \Delta -1 ) \ldots (- 2 \Delta - p) \left( (-y_0)^{-2\Delta - p +1}y_0^{\Delta - ik} + (-y_0)^{2\Delta + p -1}y_0^{- \Delta + ik} \right) \right] \nonumber\\
\end{eqnarray}
The corresponding integrals are somewhat tedious to work out in full generality. So, let us take an example: $\Delta = 1/4$, $p=-1$. This implies: $\omega_{\rm E} = 1/4$, so this cannot describe a regular Euclidean theory. In this case, the total integral evaluates to:
\begin{eqnarray}
G_{\cO}^{\rm E}  \sim -\frac{16 \left(\sqrt{\pi } \Gamma \left(\frac{1}{4}-i k\right)+2 \pi  \Gamma \left(\frac{3}{4}-i k\right)\right)}{(-4 k+i)^2 \Gamma \left(-i k-\frac{1}{4}\right)} \ . \label{corr1}
\end{eqnarray}
Possible poles can come from the Gamma functions in the numerator and the factor in the denominator. On the other hand, zeroes of the correlator can come from the pole of the Gamma function in the denominator.

\begin{figure}[ht!]
\begin{center}
\includegraphics[height=4in,width=6in,angle=0]{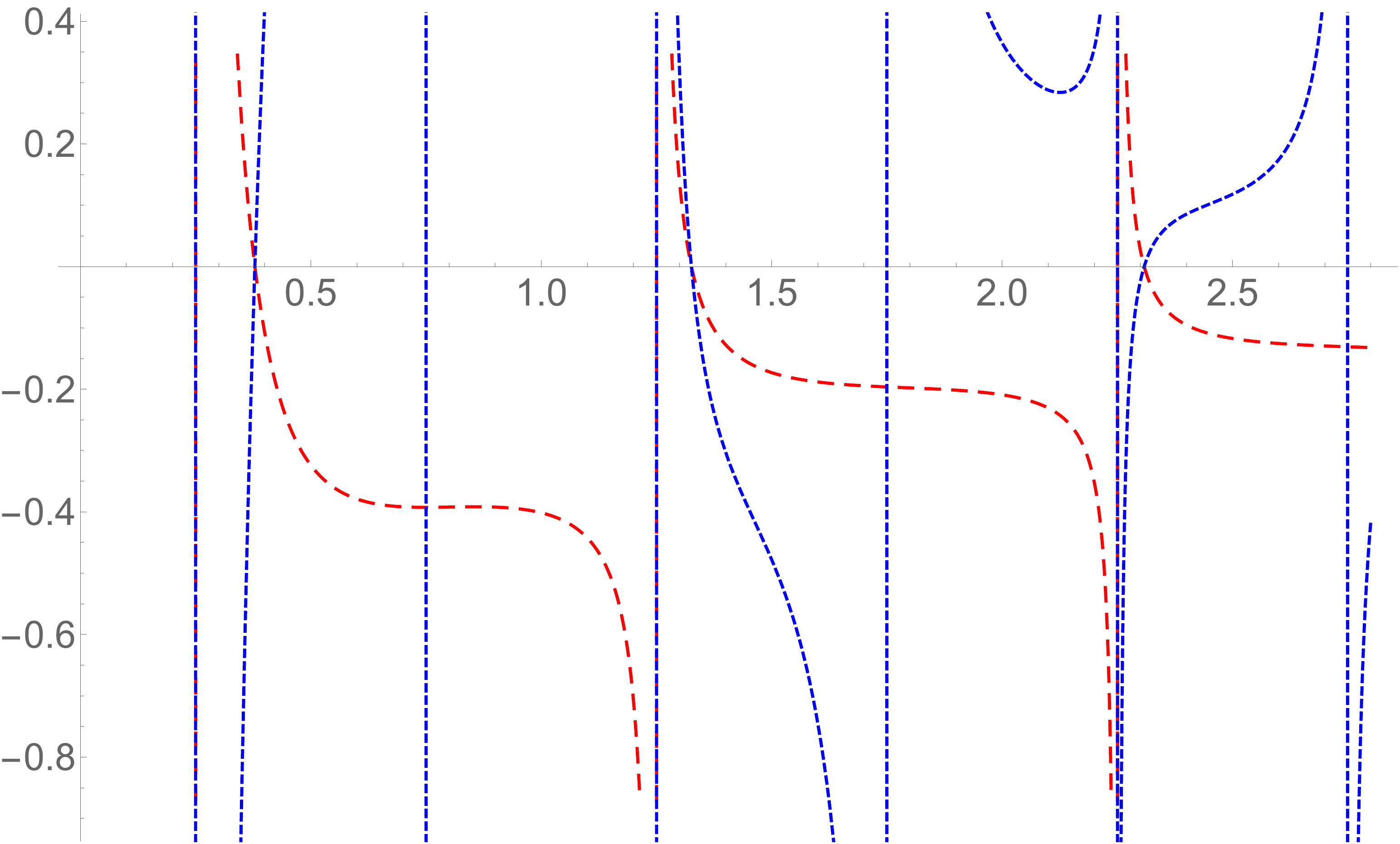}
\caption{This corresponds to $\Delta = 1/4$, $p=-1$. The red dashed line represents (\ref{corr1}) and the blue dashed line represents (\ref{corr1}) multiplied by $\Gamma \left(-i k-\frac{1}{4}\right)$. Clearly, there are additional poles in (\ref{corr1}), represented by the stand-alone blue vertical lines at $0.75$, $1.75$, {\it etc}. Thus, pole-skipping is observed in this case as well. }\label{fig1}
\end{center}
\end{figure}

\begin{figure}[ht!]
\begin{center}
\includegraphics[height=4in,width=6in,angle=0]{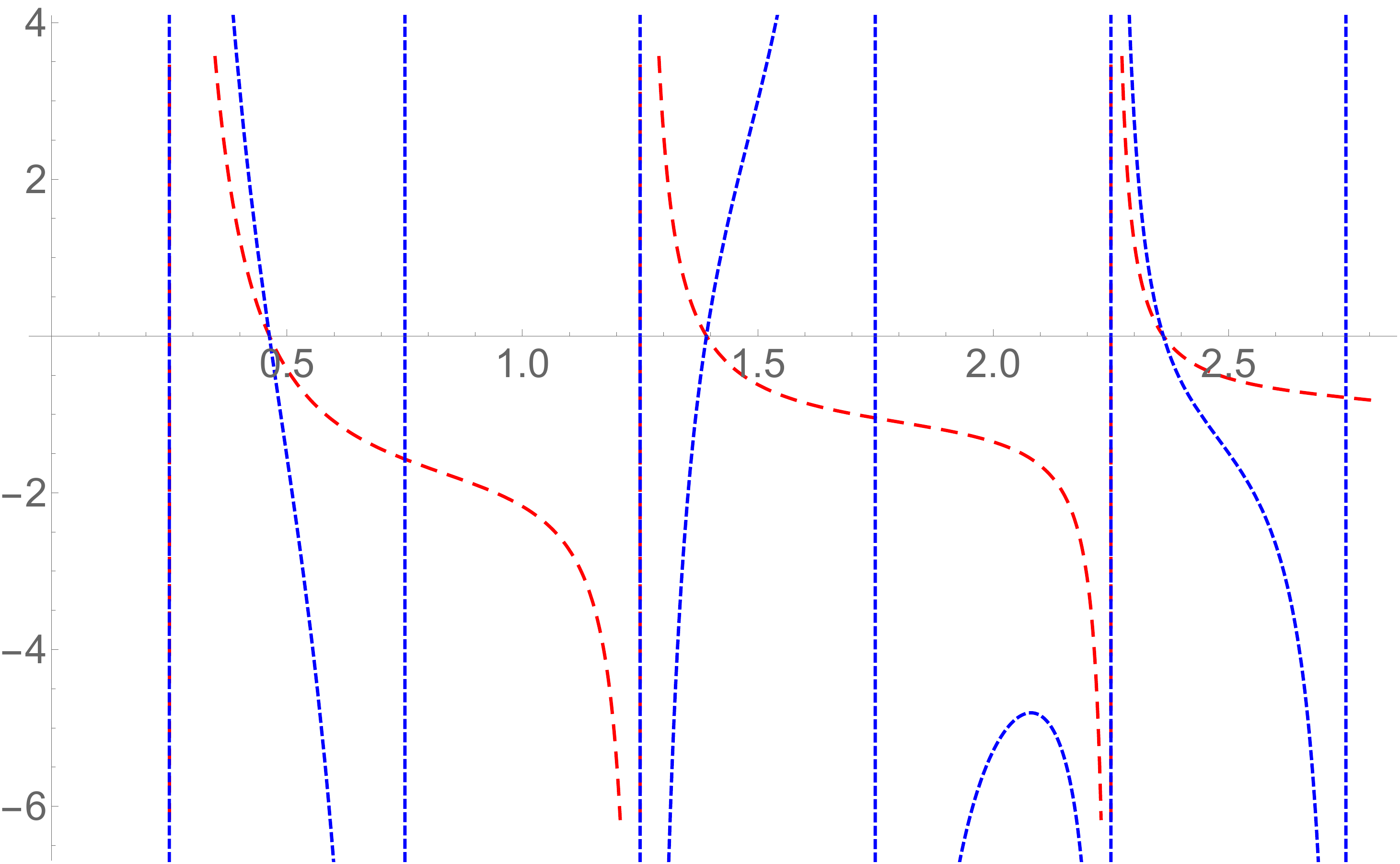}
\caption{This corresponds to $\Delta = 1/4$, $p=-2$. The red dashed line represents (\ref{corr2}) and the blue dashed line represents (\ref{corr2}) multiplied by $\Gamma \left(\frac{3}{4}-i k\right)$. Clearly, there are additional poles in (\ref{corr2}), represented by the stand-alone blue vertical lines at $0.75$, $1.75$, {\it etc}. Thus, pole-skipping is observed in this case as well..}\label{fig2}
\end{center}
\end{figure}

Similarly, consider the example of $\Delta = 1/4$, $p=-2$ ({\it i.e.}~$\omega_{\rm E} = 5/4$). This yields:
\begin{eqnarray}
G_{\cO}^{\rm E}  \sim \frac{4 i \left(\sqrt{\pi } \Gamma \left(\frac{1}{4}-i k\right)+\pi  \Gamma \left(\frac{3}{4}-i k\right)\right)}{(4 k-5 i) \Gamma \left(\frac{3}{4}-i k\right)} \ . \label{corr2}
\end{eqnarray}
The pole structure and a possible pole-skipping is easily viewed by keeping track of the pole structure of the numerator and the denominator of the expressions in (\ref{corr1}) and (\ref{corr2}). These are shown in figs.~\ref{fig1} and \ref{fig2}, respectively. 
Pictorially, it is clear that the poles of the correlators at $3/4 + n$, $n = 0 , 1, \ldots$ are subsequently skipped because of the Gamma function in the denominator. In fact, it can be shown that this remains true for any value of $p$, with given $\Delta = 1/4$. 

\newpage

(iii) $\Delta - 1- \omega_{\rm E} \in {\mathbb R} \setminus {\mathbb Z}$: This is a natural case, in which the frequencies can be quantized and a direct identification with thermal physica can be made. 
In this case, the contour is shown below. Here, the branch cut running from $0$ aligns with the branch cut running from $y_0$. We can proceed as before and replace the contour integral by its principal value. As before, we obtain:
\begin{eqnarray}
P(\I) = - \int_{l_1+l_2} - \int_{\gamma_\varepsilon} \ .
\end{eqnarray}
\begin{center}
\begin{tikzpicture}
  \tkzInit[xmin=-5,ymin=-5,xmax=5,ymax=5]

  \tkzDrawXY[noticks]
   \tkzDefPoint(0,0){O1}
\tkzDefPoint(0,0){O2}

   \tkzDefPoint(1.5,.2){B} \tkzDefPoint(1.5,-.2){C}
   \tkzDefPoint(4,.2){D} \tkzDefPoint(4,-.2){E} 
    
 \begin{scope}[decoration={markings,
      mark=at position .20 with {\arrow[scale=1]{>}};}]
  \tkzDrawArc[line width=1pt](O2,B)(C)
 \end{scope} 
  
  \begin{scope}[decoration={markings,
      mark=at position .5 with {\arrow[scale=1]{>}};}]
    \tkzDrawSegments[postaction={decorate},line width=1pt](B,D E,C)
  \end{scope}
  
  \begin{scope}[decoration={markings,
     mark=at position .20 with {\arrow[scale=1.5]{>}},
     mark=at position .80 with {\arrow[scale=1.5]{>}};
     }]
    \tkzDrawArc[postaction={decorate},line width=1pt](O1,D)(E)
  \end{scope}
  
  \draw[thick] (1,0) node[branch point,draw=red,thick] {};
   \draw[thick] (0,0) node[branch point,draw=red,thick] {};
  \draw[thick,red,branch cut] (0,0) to (0:5);
  
\node at (0.4,0.53) {$\gamma_{\varepsilon}$};
\node at (-3.4,2.8) {$\gamma_{R}$};
\node at (2.9,0.5) {$l_1$};
\node at (2.7,-0.5) {$l_2$};



\end{tikzpicture}\label{contour31}
\end{center}
Here
\begin{eqnarray}
- \int_{l_1 + l_2} & =&  \left( e^{2\pi i p} -1 \right) \int_0^1 dx x^p \frac{y_0^{\Delta - ik}}{(x - y_0)^{2\Delta}} \nonumber\\
& = & \left( e^{2\pi i p} -1 \right) y_0^{-\Delta -i k+p+1} \left(e^{2 i \pi  \Delta } B_{\frac{1}{y_0}}(p +1,1-2 \Delta )-\frac{2 i \pi  \Gamma (p+1)}{\Gamma (2 \Delta ) \Gamma (p - 2 \Delta +2)}\right) \ , \\
&& {\rm if} \quad p + 1 > 0 \ , \quad 2\Delta <1 \ , \quad 1 \ge y_0 > 0 \ . \label{cond1}
\end{eqnarray}
Here $B_{\frac{1}{y_0}}(p +1,1-2 \Delta )$ is the Beta function. Finally, 
\begin{eqnarray}
\int_{\gamma_\varepsilon} & = & \lim_{\epsilon\to 0} \int_{2\pi}^0 d\left( \epsilon e^{i\theta}\right) \epsilon^p e^{ip\theta} \frac{y_0^{\Delta - i k}}{(\epsilon e^{i\theta} - y_0)^{2\Delta}} \nonumber\\
& = & \lim_{\epsilon\to 0} \epsilon^{p+1} ({\rm finite}) \to 0 \quad {\rm for} \quad p+1 >0 \ . \label{cond2}
\end{eqnarray}
Clearly, the conditions in (\ref{cond1}) and (\ref{cond2}) have a overlapping region: $p > -1$. Away from this region, the principal value of the integral is not well-defined. The corresponding principal value is obtained to be:
\begin{eqnarray}
 \left( e^{2\pi i p} -1 \right)^{-1} P(\I) & = & y_0^{-\Delta -i k+p+1} \left(e^{2 i \pi  \Delta } B_{\frac{1}{y_0}}(p +1,1-2 \Delta )-\frac{2 i \pi  \Gamma (p+1)}{\Gamma (2 \Delta ) \Gamma (p - 2 \Delta +2)}\right) \ , \\
&& {\rm if} \quad p + 1 > 0 \ , \quad 2\Delta <1 \ , \quad 1 \ge y_0 > 0 \ . \label{cond3}
\end{eqnarray}
The final $y_0$-integral can be carried out analytically. This yields:
\begin{eqnarray}
G_{\cO}^{\rm E} \sim \frac{\frac{\Gamma (i k+\Delta ) \left(i \csc (2 \pi  \Delta ) e^{\pi  (k+i \Delta)}+e^{2 i \pi  \Delta } {\rm csch} (\pi  (k+i \Delta ))\right)}{\Gamma (i k-\Delta+1)}-\frac{e^{\pi  (k-i \Delta )} \left(2 e^{2 i \pi  (\Delta -\omega_{\rm E})}+ i\cot (2 \pi  \Delta )-1\right) \Gamma (\Delta - \omega_{\rm E} +1)}{\Gamma (-\Delta - \omega_{\rm E} + 2)}}{(k - i (\omega_{\rm E}-1)) \left( e^{2\pi i p} -1 \right)^{-1}} \ ,
\end{eqnarray}
where we have ignored overall factors that depend only on $\Delta$ and an overall phase. The first term in the numerator above purely has no dependence on $\omega_{\rm E}$ and therefore let us ignore it. The second term in the numerator also does not have any zeroes for integer valued $\omega_{\rm E}$ and real $\Delta$. Finally, let us look at the term in the denominator: $\left( e^{2\pi i p} -1 \right) $. This will yield a zero only if $p$ is an integer, which contradicts our assumption above. Thus, in this case, there is no generic pole-skipping.

We can certainly choose the two branch cuts differently. For example, consider the contour below:
\begin{center}
\begin{tikzpicture}
  \tkzInit[xmin=-5,ymin=-5,xmax=5,ymax=5]

  \tkzDrawXY[noticks]
   \tkzDefPoint(0,0){O1}
\tkzDefPoint(2,0){O2}

   \tkzDefPoint(2.5,.2){B} \tkzDefPoint(2.5,-.2){C}
   \tkzDefPoint(4,.2){D} \tkzDefPoint(4,-.2){E} 
   
   \tkzDefPoint(-0.5,.2){F} \tkzDefPoint(-0.5,-.2){G}
   \tkzDefPoint(-4,.2){H} \tkzDefPoint(-4,-.2){J} 
    
 \begin{scope}[decoration={markings,
      mark=at position .20 with {\arrow[scale=1]{>}};}]
  \tkzDrawArc[line width=1pt](O2,B)(C)
 \end{scope} 
  
  \begin{scope}[decoration={markings,
      mark=at position .5 with {\arrow[scale=1]{>}};}]
    \tkzDrawSegments[postaction={decorate},line width=1pt](B,D E,C)
  \end{scope}
  
  \begin{scope}[decoration={markings,
      mark=at position .5 with {\arrow[scale=1]{>}};}]
    \tkzDrawSegments[postaction={decorate},line width=1pt](H,F G,J)
  \end{scope}
   
  \begin{scope}[decoration={markings,
     mark=at position .80 with {\arrow[scale=1.5]{>}};
     }]
    \tkzDrawArc[postaction={decorate},line width=1pt](O1,G)(F)
  \end{scope}
  
  \begin{scope}[decoration={markings,
     mark=at position .20 with {\arrow[scale=1.5]{>}};
     }]
    \tkzDrawArc[postaction={decorate},line width=1pt](O1,D)(H)
  \end{scope}
  
   \begin{scope}[decoration={markings,
     mark=at position .20 with {\arrow[scale=1.5]{>}};
    }]
    \tkzDrawArc[postaction={decorate},line width=1pt](O1,J)(E)
  \end{scope}
  
  \draw[thick] (2,0) node[branch point,draw=red,thick] {};
   \draw[thick] (0,0) node[branch point,draw=red,thick] {};
  \draw[thick,red,branch cut] (2,0) to (0:5);
\draw[thick,red,branch cut] (0,0) to (0:-5);
  
\node at (0.5,0.6) {$\gamma_{\varepsilon_1}$};
\node at (1.5,0.6) {$\gamma_{\varepsilon_2}$};
\node at (-3.4,2.8) {$\gamma_{R}$};
\node at (2.9,0.5) {$l_1$};
\node at (2.7,-0.5) {$l_2$};
\node at (-2.9,0.5) {$l_3$};
\node at (-2.7,-0.5) {$l_4$};



\end{tikzpicture}\label{contour31}
\end{center}
On this contour, the principal value of the integral is given by
\begin{eqnarray}
P(\I) = - \int_{l_1+l_2+l_3+l_4} - \int_{\gamma_{\varepsilon_1} + \gamma_{\varepsilon_2}} \ .
\end{eqnarray}
Here
\begin{eqnarray}
\int_{\gamma_{\varepsilon_1}}  & = & \lim_{\epsilon\to 0} \int_\pi^\pi d(\epsilon e^{i\phi}) (\epsilon e^{i\phi})^p \, \frac{y_0^{\Delta - i k}}{(\epsilon e^{i\phi} - y_0)^{2\Delta}}  \nonumber\\
& = &\lim_{\epsilon\to 0} \epsilon^{1+p} ({\rm finite}) \to 0 \quad {\rm for} \quad 1+p > 0 \ . \label{cond4}
\end{eqnarray}
Similarly, 
\begin{eqnarray}
\int_{\gamma_{\varepsilon_2}}  & = & \lim_{\epsilon\to 0} \int_{2\pi}^0 d(\epsilon e^{i\theta}) (\epsilon + y_0)^p \, \frac{y_0^{\Delta - i k}}{\epsilon^{2\Delta} e^{2i \Delta\phi}}  \nonumber\\
& = &\lim_{\epsilon\to 0} \epsilon^{1-2\Delta} ({\rm finite}) \to 0 \quad {\rm for} \quad 1 > 2 \Delta \ .  \label{cond5}
\end{eqnarray}
The conditions in (\ref{cond4}) and (\ref{cond5}) readily yield the condition in (\ref{cond3}). Now, we can evaluate the integrals above and below the two branch cuts, as follows:
\begin{eqnarray}
&& \int_{l_1+l_2} = \left( 1 - e^{- 4\pi i \Delta}\right)  \int_{y_0}^1 dx (x + y_0)^p \frac{y_0^{\Delta - i k}}{x^{2\Delta}} \ , \\
&& \int_{l_3+l_4} = e^{ip\pi} \left( 1 - e^{- 2p \pi i }\right)  \int_{0}^1 dx x^p \frac{y_0^{\Delta - i k}}{(x - y_0)^{2\Delta}} \ .
\end{eqnarray}
Thus, we get
\begin{eqnarray}
P(\I) = \left( e^{- 4\pi i \Delta} - 1\right)  \int_{0}^{1-y_0} dx (x + y_0)^p \frac{y_0^{\Delta - i k}}{x^{2\Delta}} + e^{ip\pi} \left( e^{- 2p \pi i } - 1 \right)  \int_{0}^1 dx x^p \frac{y_0^{\Delta - i k}}{(x - y_0)^{2\Delta}} \ . 
\end{eqnarray}
The rest of the analyses is similar as before. To ensure a pole skipping, the numerators of both the terms above need to vanish separately. This can happen only if $2\Delta$ and $p$ are both integers, which contradicts our assumption. Thus, pole-skipping does not occur here.

\section{Retarded Correlator in BCFT: Analytic Structure}

Let us begin with the map from cylinder to UHP. We have:
\begin{align}
z_{0} = i\frac{1+\omega_{0}}{1-\omega_{0}} \ ,  \quad \omega_{0} = e^{-|x|}  \ , \\
\bar{z}_{0} = -i\frac{1+\bar{\omega}_{0}}{1-\bar{\omega}_{0}} \ ,  \quad \bar{\omega}_{0} = e^{-|x|} \ , \\
z_{1} = i\frac{1+\omega_{1}}{1-\omega_{1}} \ ,  \quad \omega_{1} = e^{i\tau}  \ . 
\end{align}
Under this map the holomorphic three point function in the full plane is given by
\begin{align}
\braket {\mathcal{O}(z_{0}) \mathcal{O}(\bar{z}_{0}) \mathcal{O}(z_{1})} &= \left(\frac{\partial z_{0}}{\partial \omega_{0}}\right)^{\Delta_{1}} \left(\frac{\partial \bar{z}_{0}}{\partial \bar{\omega}_{0}}\right)^{\Delta_{1}}\left(\frac{\partial z_{1}}{\partial \omega_{1}}\right)^{\Delta_{2}}\frac{1}{(z_{0}-\bar{z}_{0})^{2\Delta_{1}-\Delta_{2}}}\frac{1}{(\bar{z}_{0}-z_{1})^{-\Delta_{2}}}\frac{1}{(z_{1}-z_{0})^{-\Delta_{2}}} \nonumber \\
&= (i)^{\Delta_{2}} \left[\frac{2}{(1-\omega_{0})^{2}}\right]^{\Delta_{1}} \left[\frac{2}{(1-\bar{\omega}_{0})^{2}}\right]^{\Delta_{1}}\left[\frac{2}{(1-\omega_{1})^{2}}\right]^{\Delta_{2}} \left[\frac{2i(1-\omega_{0}\bar{\omega}_{0})}{(1-\omega_{0})(1-\bar{\omega}_{0})}\right]^{\Delta_{2}-2\Delta_{1}} \nonumber \\
& \times \left[\frac{-2i(1-\omega_{1}\bar{\omega}_{0})}{(1-\omega_{1})(1-\bar{\omega}_{0})}\right]^{-\Delta_{2}} \left[\frac{2i(\omega_{1}-\omega_{0})}{(1-\omega_{1})(1-\omega_{0})}\right]^{-\Delta_{2}} \nonumber \\
& = (i)^{2(\Delta_{2}-\Delta_{1})} (1-\omega_{0}\bar{\omega}_{0})^{\Delta_{2}-2\Delta_{1}} (1-\omega_{1}\bar{\omega}_{0})^{-\Delta_{2}}(\omega_{1}-\omega_{0})^{-\Delta_{2}} \nonumber \\
& = (-1)^{\Delta_{2}-\Delta_{1}} (1-e^{-2|x|})^{\Delta_{2}-2\Delta_{1}} (1-e^{i\tau-|x|})^{-\Delta_{2}}(e^{i\tau}-e^{-|x|})^{-\Delta_{2}}
\end{align}
The Fourier transform of this Euclidean correlator is given by
\begin{align}
G(\omega,k) = \int^{\infty}_{0} dx \int^{2\pi}_{0}d\tau \frac{e^{-ikx}e^{-i\omega \tau}}{(1-e^{-2x})^{-\Delta_{2}+2\Delta_{1}} (1-e^{i
\tau-x})^{\Delta_{2}}(e^{i\tau}-e^{-x})^{\Delta_{2}}}  \ .
\end{align}
By denoting $z=e^{i\tau}$ and by taking $\Delta_{2} = 2$(i.e taking boundary operators $\mathcal{O}$ to be holomorphic stress energy tensor $T(z)$), we get:
\begin{align}\label{main bcft}
G(\omega,k) &= -(-1)^{-\Delta_{1}}i \int^{\infty}_{0}dx \frac{e^{-ikx}}{\left(1-e^{-2x}\right)^{2\Delta_{1}-2}}\oint_{|z|=1}\frac{dz}{z^{\omega+1}\left(1-ze^{-x}\right)^{2}\left(z-e^{-x}\right)^{2}} \nonumber \\
& = -(-1)^{-\Delta_{1}}i \int^{\infty}_{0}dx \frac{e^{-ikx+2x}}{\left(1-e^{-2x}\right)^{2\Delta_{1}-2}}\oint_{|z|=1}\frac{dz}{z^{\omega+1}\left(z-e^{x}\right)^{2}\left(z-e^{-x}\right)^{2}} \ . 
\end{align}
The $z$ integral over the unit circle, has poles at $z=0$ and $z=e^{-x}$ for the given integration range of $x$. However, the pole at $z=0$ comes only from the definite set of $\omega$ when $\omega > -1$. By considering this restriction on $\omega$, we can evaluate the residues coming from $z$ integral. It yields:
\begin{align}
\oint_{|z|=1}\frac{dz}{z^{\omega+1}\left(z-e^{x}\right)^{2}\left(z-e^{-x}\right)^{2}} = 2\pi i\left(I_{1}(z=0)+I_{2}(z=e^{-x})\right) \ ,
\end{align}
where
\begin{align}
&I_{1} = \frac{\left((\omega+3)e^{-(\omega+1)x}-(\omega+3)e^{(\omega+1)x}+(\omega+1)e^{(\omega+3)x}- (\omega+1)e^{-(\omega+3)x}\right)}{\left(e^{x}-e^{-x}\right)^{3}}; \quad \text{and} \\
&I_{2} = -\frac{\left((\omega+1)e^{(\omega+2)x}(e^{-x}-e^{x})+2e^{(\omega+1)x}\right)}{\left(e^{-x}-e^{x}\right)^{3}} \ .
\end{align}
Putting it back in the expression (\ref{main bcft}), we get
\begin{align}
G(\omega,k) &= (-1)^{-\Delta_{1}}2\pi \bigg( (\omega+3)\int^{\infty}_{0}dx \frac{e^{-ikx-x-(\omega+1)x}}{\left(1-e^{-2x}\right)^{2\Delta_{1}+1}} - (\omega+3)\int^{\infty}_{0}dx \frac{e^{-ikx-x+(\omega+1)x}}{\left(1-e^{-2x}\right)^{2\Delta_{1}+1}} \nonumber \\
& + (\omega+1)\int^{\infty}_{0}dx \frac{e^{-ikx-x+(\omega+3)x}}{\left(1-e^{-2x}\right)^{2\Delta_{1}+1}} - (\omega+1)\int^{\infty}_{0}dx \frac{e^{-ikx-x-(\omega+3)x}}{\left(1-e^{-2x}\right)^{2\Delta_{1}+1}} \bigg)_{I_{1}} \nonumber \\
& - (-1)^{-\Delta_{1}}2\pi \bigg( (\omega+1)\int^{\infty}_{0}dx \frac{e^{(\omega - ik)x}}{\left(1-e^{-2x}\right)^{2\Delta_{1}}} + 2 \int^{\infty}_{0}dx \frac{e^{(\omega - ik)x}}{\left(1-e^{-2x}\right)^{2\Delta_{1}+1}}\bigg)_{I_{2}}
\end{align}
Using analytic continuation of Beta function, the integral identity $\int^{\infty}_{0}dx \frac{e^{ax}}{\left(1-e^{-2x}\right)^{b}} = \frac{1}{2}\frac{\Gamma(-\frac{a}{2})\Gamma(-b+1)}{\Gamma(-b-\frac{a}{2}+1)}$ yields
\begin{align}
G(\omega,k) &= (-1)^{-\Delta_{1}} \bigg( \pi (\omega+3) \frac{\Gamma\left(\frac{\omega+ik}{2}+1\right)\Gamma\left(-2\Delta_{1}\right)}{\Gamma\left(-2\Delta_{1}+1+\frac{\omega+ik}{2}\right)} - \pi (\omega+3) \frac{\Gamma\left(-\frac{\omega-ik}{2}\right)\Gamma\left(-2\Delta_{1}\right)}{\Gamma\left(-2\Delta_{1}-\frac{\omega-ik}{2}\right)}  \nonumber \\
& + \pi (\omega+1) \frac{\Gamma\left(-\frac{\omega-ik}{2}-1\right)\Gamma\left(-2\Delta_{1}\right)}{\Gamma\left(-2\Delta_{1}-1-\frac{\omega-ik}{2}\right)} - \pi (\omega+1) \frac{\Gamma\left(\frac{\omega+ik}{2}+2\right)\Gamma\left(-2\Delta_{1}\right)}{\Gamma\left(-2\Delta_{1}+2+\frac{\omega+ik}{2}\right)} \nonumber \\
& - \pi (\omega+1) \frac{\Gamma\left(-\frac{\omega-ik}{2}\right)\Gamma\left(-2\Delta_{1}+1\right)}{\Gamma\left(-2\Delta_{1}+1-\frac{\omega-ik}{2}\right)} -2\pi \frac{\Gamma\left(-\frac{\omega-ik}{2}\right)\Gamma\left(-2\Delta_{1}\right)}{\Gamma\left(-2\Delta_{1}-\frac{\omega-ik}{2}\right)}\bigg) \ . 
\end{align}

{\large\bf Analysis of pole skipping(for $\omega > -1$)}

Thus the propagator in the momentum space, contains six terms having simple poles coming from numerators of Gamma functions with non-positive integer values. We will now analyze each six terms to see where pole skipping occurs.
\begin{itemize}
\item The first term containing simple poles coming from the numerator, has the structure $\frac{\omega+ik}{2}+1 = -n$, $n \in \mathbb{Z}_{+}$. Hence the poles lies on the $\omega = -2(n+1) - ik$ line. These poles could be cancelled from that of the Gamma function of the denominator where the poles lies on the line $\omega = -2(m-2\Delta_{1}-1)-ik$, where $m \in \mathbb{Z}_{+}$. Hence the pole skipping occurs for $\Delta_{1} = \frac{(m-n)-2}{2}$ and it implies $m> n+2$. But since $\omega >-1$, no pole skipping occurs from this term. 
\item A similar analysis could show that pole skipping can happen from the second,third,fifth and sixth term at $\omega -ik = n$($n=1,2,\dots$),$\omega -ik = n$($n=0,1,\dots$), $\omega -ik = n$($n=1,2,\dots$), $\omega -ik = n$($n=1,2,\dots$) for any $\Delta_{1} \in \mathbb{R}_{+}$. Like the first term, the fourth term also does not show any pole skipping. Hence, as expected, the pole skipping is observed at discrete Matsubara frequiencies and $\omega = 1$ is identified with the Lyapunov exponent as being maximal. 
\item One could also show that for $\omega <-1$, there will be no pole skipping from the terms contributing in this case, i.e the fifth and the sixth term.
\end{itemize}

A similar analysis could be carried out by taking $\Delta_{2} = 1$ and one could get the final Fourier transform of the correlator $G(\omega,k)$ as (for $\omega>-1$)
\begin{align}
G(\omega,k) \sim \frac{\Gamma\left(\frac{\omega + ik +1}{2}\right)\Gamma\left(-2\Delta_{1}+1\right)}{\Gamma\left(-2\Delta_{1}+1+\frac{\omega + ik +1}{2}\right)} \ . 
\end{align}
It is straightforward to check that no pole skipping occurs here as $\omega > -1$. It would be interesting to illuminate similar analysis for general $\Delta_{2}$.

\section{An $n$-point Correlator with a Boundary}

In two dimensions, conformal transformations are holomorphic transformations of the complex coordinates : $(z,\bar{z})\rightarrow (\omega(z), \bar{\omega}(\bar{z}))$. The infinitesimal version maybe expressed as $z' = z+\epsilon(z)$, $\bar{z}' =\bar{z} +\bar{\epsilon}(\bar{z})$. A two dimensional conformal field theory is invariant under independent transformations in the $z$ and $\bar{z}$ coordinates, $ie$ with independent $\epsilon$ and $\bar{\epsilon}$. In the radially quantized  CFT, these transformations are generated by $\oint_{|z|=1} T(z) \epsilon(z) dz$ and $\oint_{|z|=1} \bar{T}(\bar{z})\bar{\epsilon}(\bar{z})d\bar{z}$,  respectively.  The variation of a n-point correlation function of primary fields under an arbitary infintesimal conformal is given by the sum of the two independent variations of the holomorphic and anti-holomorphic part:
\begin{eqnarray}\label{nptcft}
\delta_{\epsilon ,\bar{\epsilon}}\langle\phi_1(z_1,\bar{z}_1)....\phi_n(z_n,\bar{z}_n)\rangle = && \sum^n_{i=1} \langle\phi_1(z_1,\bar{z}_{1})...\delta_{\epsilon_i}(\phi_i(z_i,\bar{z}_i))...\phi_n(z_n,\bar{z}_n)\rangle \nonumber \\
&& +\sum^{n}_{i=1}\langle\phi_1(z_1,\bar{z}_{1})...\delta_{\bar{\epsilon}_i}(\phi_i(z_i,\bar{z}_i))...\phi_n(z_n,\bar{z}_n)\rangle  \ .
\end{eqnarray} 

Now, if we consider a BCFT with boundary given by $z=\bar{z}$, we only have the subset of symmetry transformations which preserve the boundary. $ie:$ $\left. \Big[z +\epsilon(z) = \bar{z} +\bar{\epsilon}(\bar{z})\Big] \right |_{z=\bar{z}}$. Therefore the boundary preserving conformal transformations are given by $\bar{\epsilon}(\bar{z}) = \epsilon(\bar{z})$, so that we only have one set of independent conformal transformations. Thus we can express the variation of the n-pt correlation function in this case as follows:  

\begin{equation}\label{nptBCFT}
\delta_{\epsilon ,\bar{\epsilon}}\langle\phi_1(z_1,\bar{z}_1)....\phi_n(z_n,\bar{z}_n)\rangle = \sum^{2n}_{I=1} \langle\phi_1(z_1,\bar{z}_{1})...\delta_{\epsilon_I}(\phi(z_i,\bar{z}_i))...\phi_n(z_n,\bar{z}_n)\rangle  \ .
\end{equation}

It is clear then that the variation of an n-point correlation function in the BCFT now has the same form as the purely holomorphic tranformation of a 2n-pt correlation function, of 2n fields $\phi_{I}$ with holomorphic coordinates $z_{I}$, with $I$ running from $I=1...2n$. The weights of the fields given by $h_I$, where $h_I= h_i$ for $I=1,..n$ and $h_I =\bar{h}_i$ for the remaining set of fields.  These fields are located at the positions $z_{I}=z_i$ for $I=1,...n$ and $z_I =\bar{z}_i$ for $I=n+1,...2n$, respectively.

In particular a two point function in a BCFT has the same transformation properties as a purely holomorphic four point function of a CFT, with no boundaries. These conformal transformation properties, fixes the form of the four point function upto a undetermined function of the cross ratio: 
\begin{equation}
\langle \phi_1(z_1)\phi_2(z_2)\phi_3(z_3)\phi_4(z_4)\rangle = \Big(\prod^4_{i<j}z^{h/3-h_i-h_j}_{ij}\Big)F(z) \; \; \textrm{where}\; \;  z=\frac{z_{12}z_{34}}{z_{13}z_{24}}  \textrm{and}\; \;  h=\sum^4_{i=1} h_i
\end{equation}
Thus a two point function in a BCFT will be of the above form, with $z_2 =\bar{z}_{1}$ and $z_4 =\bar{z}_2$ and the conformal dimensions of the fields of the operators at $z_3$ and $z_4$ will be $h_3=\bar{h}_{1}$ and $h_4=\bar{h}_2$ respectively.

\section{Kubo-Martin-Schwinger Condition}

The so-called KMS condition can be viewed as the definition of an equilibrium state. Given a finite dimensional Hilbert space, a trace class operator $\rho = e^{-\beta H}$ (which is the density matrix), and a set of self-adjoint operators $\{\cO_i\}$, where $i$ runs over all possible observables, the KMS condition on the $2$-point correlation function is given by
\begin{eqnarray}
{\rm Tr} \left( e^{-\beta H} \cO_1(t) \cO_2(0) \right) \equiv \langle \cO_1(t) \cO_2(0) \rangle_\beta =  \langle \cO_2(0) \cO_1(t + i\beta)\rangle_\beta  \ ,
\end{eqnarray}
where $\beta$ is the period of the Euclidean thermal circle. The above condition can be easily obtained by noting:
\begin{eqnarray}
\cO_1(t) = e^{i H t} \cO_1(0) e^{- i H t} \ , 
\end{eqnarray}
where $H$ is the corresponding Hamiltonian and the cyclic property of trace.

It is reasonable to assume that all operators in the system acquire at most a phase under $t \to t + i\beta$, since the Euclidean time is periodic and one needs to impose boundary conditions on the corresponding spectrum. For example, for a Bosonic operator, we expect $\cO(t + i\beta) = \cO(t)$ and for a Fermionic one $\cO(t+ i\beta) = - \cO(t)$. In general, we can allow a boundary condition $\cO(t+ i\beta) = e^{i\alpha} \cO(t)$, where $\alpha \in {\mathbb R}$. The upshot is: with the reversed operator ordering, the $2$-point correlator is proportional to the original time-ordered correlator.

Let us now use the same condition on a $3$-point correlator. Consider the following correlator:
\begin{eqnarray}
\langle \cO_1(t_1) \cO_2(t_2) \cO_3(0) \rangle_\beta = \langle \cO_3(0) \cO_1(t_1+i\beta) \cO_2(t_2+i\beta) \rangle_\beta \ ,
\end{eqnarray}
with the following time-ordering: $t_2 > t_1 > 0$. Manifestly, the LHS is an OTOC, which, upon using the KMS condition reduces to a TOC. It is straightforward to check that, for arbitrary ordering of time arguments, a $3$-point correlator can always be reduced to a time-ordered one. On the other hand, $4$-point and higher correlators cannot be reduced to correlators with a monotonic time-ordering, generically.

The above arguments make minimal assumptions. The presence of a non-trivial boundary, in fact, affects the state in which the expectation value is measured. For a CFT, this is particularly simple: One needs to simply consider a thermal state in the presence of a non-trivial boundary state itself. Let us briefly review the notion of boundary states here, in $2$D.

The physical requirement on a CFT with a boundary is that there is no source or sink of energy momentum at the boundary. In terms of the Virasoro generators, this implies:
\begin{eqnarray}
\left( L_n - \bar{L}_{-n} \right)  | b \rangle = 0  \ ,
\end{eqnarray}
where $| b \rangle$ is the boundary state and $L_n$, $\bar{L}_{-n}$ are Virasoro generators on the plane. In a given Verma module ${\cal V}_j \otimes \bar{{\cal V}}_j$, the above condition is satisfied by the so-called Ishibashi states, defined as:
\begin{eqnarray}
| j \rangle \rangle \equiv \sum_N | j, N \rangle \otimes | \bar{j}, N \rangle \ ,
\end{eqnarray}
with $j = \bar{j}$. Here $| j, N \rangle$ correspond to the descendant states at level $N$ in the given Verma module. Given the Ishibashi states, one can now define a Cardy state, which corresponds to a physical boundary state:
\begin{eqnarray}
| B \rangle \equiv \sum_j C^j | j \rangle \rangle \ ,
\end{eqnarray}
where $C^j$ are some coefficients. By construction, the Ishibashi states and therefore the Cardy states are invariant under $\left( L_0 - \bar{L}_0 \right) $ (this corresponds to rotation in the $z$-plane), but not necessarily under $\left( L_0 + \bar{L}_0 \right)$ (this corresponds to dilatation in the $z$-plane). The dilatation transformation generates time-evolution, and therefore generic boundary states may evolve in time, see {\it e.g.}~\cite{Guo:2017rji}. It is, therefore, unsurprising that KMS conditions are subtle for BCFT systems.

Furthermore, the Cardy state, generically, is not normalizable. To construct a normalizable boundary state, one considers:
\begin{eqnarray}
| B\rangle _{\ell} \equiv e^{- \ell H} | B\rangle \ ,
\end{eqnarray}
where $\ell$ is a regularization parameter hitherto undetermined. Typically, one sets $\ell = \beta/4$. The state $|B\rangle$ can itself be explicitly constructed in some cases. For instance, in a free massless scalar field theory, it takes the form $|B\rangle = \hat{B}| 0 \rangle$, where $\hat{B} \propto \exp(-\sum^{\infty}_{n=1}\frac{1}{n}a_{-n}\tilde{a}_{-n})$. The $a_n$ and $\tilde{a}_n$ being the creation operators of the chiral and anti-chiral sectors respectively. In general, ofcourse the form of $\hat{B}$ is not known. 

On such states, the correlator takes the schematic form:
\begin{eqnarray}
{\rm Tr} \left( e^{-\beta H} e^{-\ell H}\hat{B} \cO_1(t_1) \cO_2(t_2) \cO_3(0) \right ) \ , 
\end{eqnarray}
which does not have a trivial reduction in terms of a TOC. In fact, it appears that the correlators of the above kind may satisfy a different KMS-type condition in the presence of a boundary. We will not explore this any further in this article.

\end{document}